\documentclass[aps,prl,twocolumn, superscriptaddress]{revtex4-1}
\usepackage{graphicx,color}
\usepackage{amsmath,amssymb,latexsym}
\usepackage{mathtools}
\usepackage[english]{babel}
\usepackage{dcolumn}
\usepackage{bm}
\usepackage{float}
\usepackage[normalem]{ulem}
\usepackage[usenames,dvipsnames]{xcolor}

\usepackage{color}

\ProvideTextCommand{\DJ}{OT1}{\leavevmode\raisebox{-.5ex}{\makebox[0pt][l]{\hskip-.07em\accent"16\hss}}D}

\begin{document}

\title{Being heterogeneous is advantageous: Extreme Brownian non-Gaussian searches}

\author{Vittoria Sposini}
\email{vittoria.sposini@univie.ac.at}
\affiliation{
  Faculty of Physics, University of Vienna, Kolingasse 14-16, 1090 Vienna, Austria
}

\author{Sankaran Nampoothiri}
\email{sparampo@gitam.edu}
\affiliation{
Department of Physics, Gandhi Institute of Technology and Management (GITAM) University, Bengaluru-561203, India
}

\author{Aleksei Chechkin}
\email{chechkin@uni-potsdam.de}
\affiliation{
  Faculty of Pure and Applied Mathematics, Hugo Steinhaus Center, Wroclaw University of Science and Technology, Wyspianskiego Str. 27, 50-370 Wroclaw, Poland
}
\affiliation{
  Institute for Physics \& Astronomy, University of Potsdam, 14476 Potsdam-Golm,
  Germany
}
\affiliation{
  Akhiezer Institute for Theoretical Physics, 61108 Kharkov, Ukraine
}

\author{Enzo Orlandini}
\email{enzo.orlandini@unipd.it}
\affiliation{
Dipartimento di Fisica e Astronomia `G. Galilei' - DFA, Sezione INFN,
Universit\`a di Padova,
Via Marzolo 8, 35131 Padova (PD), Italy
}

\author{Flavio Seno}
\email{flavio.seno@unipd.it}
\affiliation{
Dipartimento di Fisica e Astronomia `G. Galilei' - DFA, Sezione INFN,
Universit\`a di Padova,
Via Marzolo 8, 35131 Padova (PD), Italy
}

\author{Fulvio Baldovin}
\email{fulvio.baldovin@unipd.it}
\affiliation{
Dipartimento di Fisica e Astronomia `G. Galilei' - DFA, Sezione INFN,
Universit\`a di Padova,
Via Marzolo 8, 35131 Padova (PD), Italy
}

\date{\today}

\begin{abstract}
  Redundancy in biology may be explained by the need to optimize
  extreme searching processes, where one or few among many particles are
  requested to reach the target like in human fertilization.
  We show that non-Gaussian rare fluctuations in
  Brownian diffusion dominates such searches, introducing drastic
  corrections to the known Gaussian behavior.
  Our demonstration
  entails different physical systems
  and pinpoints the relevance of diversity within
  redundancy to boost fast targeting.
  We sketch an experimental context to test our results: polydisperse
  systems.
\end{abstract}

\maketitle

With the discovery~\cite{wang2009} of Brownian non-Gaussian (BnG) diffusion 
-- a stochastic motion with a mean squared displacement linearly increasing 
in time (Brownian or Fickian behavior) and a non-Gaussian probability density 
function (PDF) for the displacements -- the expectation has been raised~\cite{wang2012} 
that the excess of probability for rare large fluctuations might dominate 
first-passage processes, leading to unexpected phenomena.
While BnG behavior 
found numerous experimental~\cite{wang2012,toyota2011,yu2013,yu2014,chakraborty2020,
weeks2000,wagner2017,jeon2016,yamamoto2017,stylianidou2014,parry2014,munder2016,
cherstvy2018,li2019,cuetos2018,hapca2008,pastore2021rapid,giavazzi2022,dean2023} and 
molecular dynamics~\cite{pastore2015,miotto2021length,pastore2022} confirmations, 
at odds with expectation theoretical analyses showed that typical Gaussian searches 
turn out to be more effective than non-Gaussian ones~\cite{grebenkov2018,grebenkov2019,
sposini2019first}. In a companion paper~\cite{sposini2023pre}, where full references 
about the available theoretical models for BnG diffusion are provided, we give a 
detailed account of this basic issue, showing that for the large class of subordination 
processes~\cite{Feller1968,bochner2020harmonic} the typical time scale for one 
searcher to reach the target -- e.g. the mean first passage time (MFPT) -- is indeed 
shorter in Gaussian than in non-Gaussian motion.

In the last years, however, an upsurge of studies and 
commentaries~\cite{redner2015,holcman2019a,holcman2019b,redner2019,martyushev2019,
rusakov2019,sokolov2019,coombs2019,tamm2019,holcman2019c,lawley2020a,lawley2020b} 
has pointed out that in many situations such as fast activation processes in chemistry 
and cellular responses in biology, the relevant timescale is actually not the time 
spent by a given single searcher to reach the target, but rather the time at which 
the first few searchers, out of many,  perform this task. 
A paradigmatic example is human reproduction, in which a single sperm cell out 
of $M\sim 10^8$ finds and fertilizes the egg. The computation of this time scale 
is a typical extreme statistics problem that justifies  the presence of redundancy 
in some biological processes but that, so far, has been mainly studied for normal 
Brownian motion. 

In this Letter, we investigate the role that BnG motion, and in general
the class of subordination processes may have on the extreme targeting
problems. In particular, 
by focusing on the diffusing
diffusivity (DD) model~\cite{chechkin2017}, and polydisperse polymer ensembles --
equilibrium grand canonical~\cite{deGennes1979,vanderzande1998} and quenched~\cite{flory1953} -- we show
that the \textit{extreme}-MFPT for
non-Gaussian diffusion may become orders of magnitude shorter than the
Gaussian one. This finding reveals a drastically different scenario
with respect to the ordinary MFPT problem, identifying extreme
targeting as a natural setting in which the non-Gaussianity makes a
substantial difference.

\begin{figure}[t]
  \includegraphics[width=0.95\columnwidth]{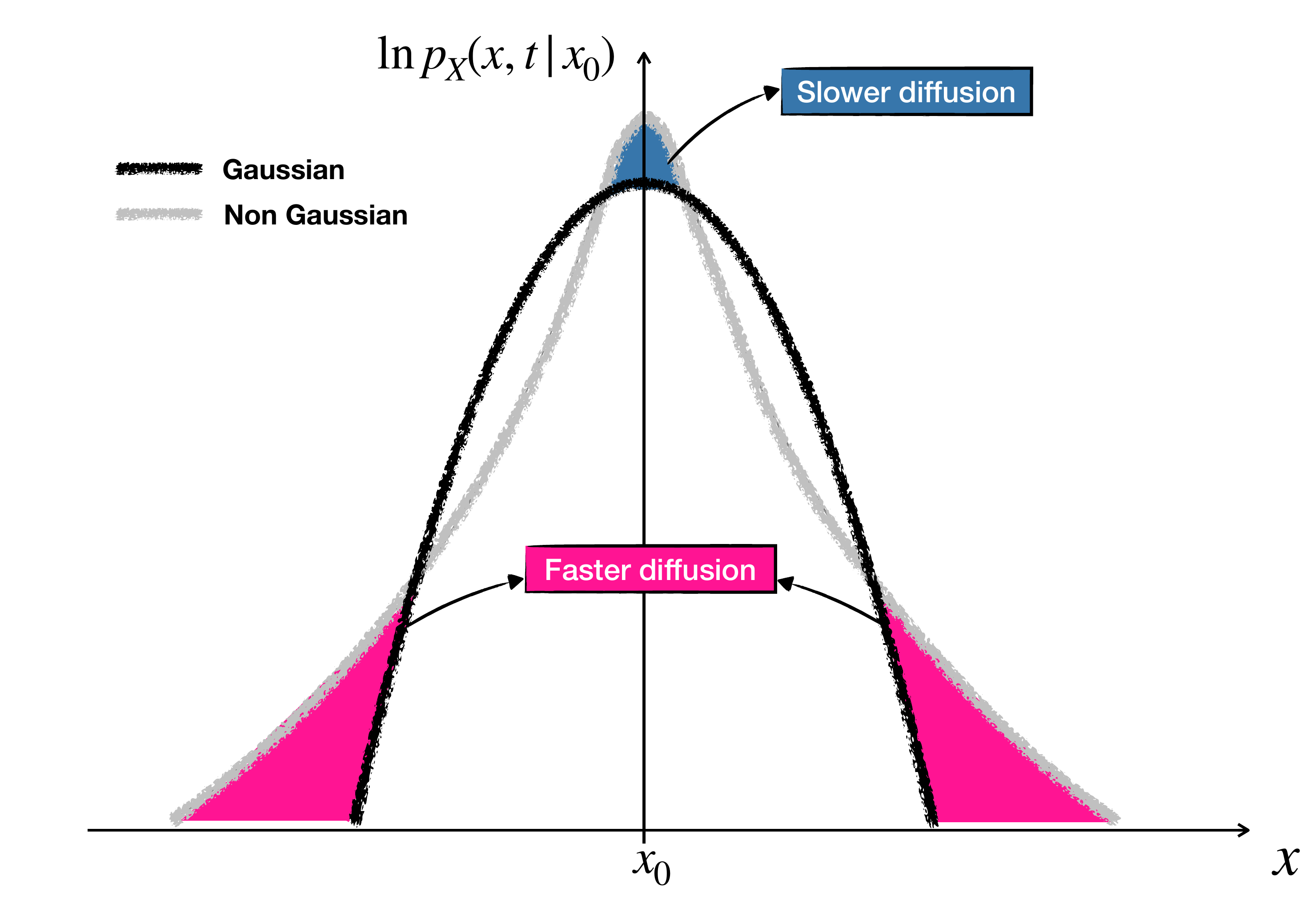}\\
  \caption{Comparison between Gaussian and non-Gaussian PDFs for
    subordination processes. The two PDFs share the same mean and standard
    deviation but the non-Gaussian one has an excess probability
    both in the tails and in the center part.
    The non-Gaussian PDF is obtained from the FSP model with $p=0.99$ (See text).
    }
  \label{fig_excess_pdf}
\end{figure}

Before going into the details of the calculations let us provide a qualitative argument for our findings. As we articulate in
Ref.~\cite{sposini2023pre},
models for BnG diffusion display an excess
of probability both in the central part and in the tails, when
compared with a Gaussian PDF of the same width (see
Fig.~\ref{fig_excess_pdf}).
The excess of probability in the central part of the PDF,
associated with a slower diffusion, is shown in~\cite{sposini2023pre} to
be responsible for the lower effectiveness of non-Gaussian searches in
ordinary targeting problems, when one looks at the typical time scale
for a particle to reach the target. At a glance, this
conclusion~\cite{grebenkov2018,grebenkov2019,sposini2019first}
frustrates expectations~\cite{wang2012} of novel phenomena in
diffusion-limited reactions driven instead by the ``tail effect''.
Yet, if one considers the class of problems in which reactions are
activated by the first (or the first few) successful searchers among
many, the corresponding targeting time scale, the
\textit{extreme}-MFPT, is governed by rare trajectories which are the
few among the many to follow a quasi-geodesic path to the
target~\cite{holcman2019a,lawley2020b}. Here we argue that through the
``tail effect'' non-Gaussianity adds to these rare events the
possibility for the searcher to diffuse faster (see
Fig.~\ref{fig_excess_pdf}) and hence it impacts dramatically the
\textit{extreme}-MFPT, as we detail below.

Let us first briefly recall the theoretical context behind the
\textit{extreme}-MFPT problem. Given $M\gg 1$ independent
(i.e. non-interacting) searchers, each with its own random arrival
time $\tau_i$, the arrival time of the fastest one is defined as
$
T_{M} = \min \left [ \{ \tau_1,\tau_2,\cdots,\tau_M\}\right] 
$.
(More generally one can consider $T_{k,M}$, namely the time at which
the $k$-fastest searches have reached the
target~\cite{lawley2020a,lawley2020b};
clearly, $T_{M}\equiv T_{1,M}$.)
We now include a possible heterogeneity for the diffusing particles,
assuming that their diffusion coefficients
are characterized either by a discrete steady-state
probability mass function $p_D^*(D_n)$ ($n=1,2,\ldots$)
or continuous PDF $p_D^*(D)$, with average
$\mathbb{E}[D]\equiv D_{\mathrm{av}}$.
Since the searchers are independent, the
statistics of $T_{M}$ can be computed from the one of a single
particle.
Denoting by $0\leq P(\tau_i > t)\leq1$ the
one-particle survival probability, and by
$\mathcal{S}_{D_n}(t)$ the survival probability of a generic particle with
diffusion coefficient $D_n$,
the probability associated with the
extreme statistics is
$
  P(T_M>t)=\prod_{i=1}^MP(\tau_i > t)
  =\prod_n\left(\mathcal{S}_{D_n}(t)\right)^{M_n}
$,
where, by the law of large numbers, $M_n=M\;p_D^*(D_n)$ is the number of
searchers with diffusion coefficient $D_n$.
The \textit{extreme}-MFPT,
$\mathbb{E}[T_M]=\int_0^\infty\mathrm{d}t\,P(T_M>t)$ is thus
\begin{equation}
  \mathbb{E}[T_M]
  =\int_0^\infty\mathrm{d}t\,\exp\left(M\sum_np_D^*(D_n)\,\ln\mathcal{S}_{D_n}(t)\right)\,.
\label{eq_extreme_mfpt}
\end{equation}
Here and below, are the substitutions
$\sum_n\mapsto\int\mathrm{d}D$,
$D_n\mapsto D$ understood if $p_D^*$ is a PDF instead of
a probability mass function.
Note that by choosing $p_D^*(D)=\delta(D-D_{\mathrm{av}})$,
Eq.~\eqref{eq_extreme_mfpt} recovers the ordinary
approach, appropriate for a Gaussian diffusion in which all particles
share the same diffusion coefficient $D_{\mathrm{av}}$.  
Despite the hypothesis of 
independent searchers enormously simplify the computation of
$P(T_M>t)$, the full explicit expression of
$\mathcal{S}_{D_n}(t)$
is often unknown and approximations are needed.  The assumption
$M\gg 1$ suggests that the computation of $P(T_M>t)$
can be approximated by the
short-time behavior of $\mathcal{S}_{D_n}(t)$, where
$\mathcal{S}_{D_n}(t)\simeq1$.
This is usually done by solving explicitly the boundary problem of the
associated Fokker-Planck equation and taking the small time
approximation of the corresponding survival
probability~\cite{yuste2000diffusion,holcman2015stochastic,holcman2019b,holcman2019a}.
For several targeting processes with varying space
dimensions, boundaries, and shape of the target (if small enough),
most results have been shown to fall into a
universal category of extreme events statistics.
This is due to the fact that the most effective rare trajectories
almost follow a geodesic path to the
target~\cite{holcman2019a,lawley2020b} of length $\ell$,
which is a straight line in a homogeneous and isotropic environment.
It is thus paradigmatic to address the one-dimensional case for which
\begin{equation}
\mathcal{S}_{D_n}(t)
=\mathrm{erf}\left(\dfrac{\ell}{\sqrt{4\,D_n\,t}}\right)
\;\substack{\simeq\\t\to0^+}\;
1-\dfrac{
  \mathrm{e}^{-\frac{\ell^2}{4\,D_n\,t}}
  \sqrt{4\,D_n\,t}}
{\sqrt{\pi}\,\ell}\,,
\end{equation}
implying
\begin{equation}
  \mathbb{E}[T_M]=
  \int_0^\infty\mathrm{d}t
  \exp\left(
  -M
  \left[\sum_np_D^*(D_n)
  \dfrac{
    \mathrm{e}^{-\frac{\ell^2}{4\,D_n\,t}}
    \sqrt{4\,D_n\,t}
  }{
    \left(\sqrt{\pi}\,\ell\right)
  }\right]
  \right).
  \label{eq_exponential}
\end{equation}
For large $M$ this integral can be approximated~\cite{redner2015} as
$\mathbb{E}[T_M]\simeq\int_0^{t_0}\mathrm{d}t=t_0$, with $t_0$
solution of
\begin{equation}
  M
  \sum_np_D^*(D_n)
  \left(
    \mathrm{e}^{-\frac{\ell^2}{4\,D_n\,t_0}}
  \middle/
    \left(\dfrac{\sqrt{\pi}\,\ell}{\sqrt{4\,D_n\,t_0}}\right)
    \right)=1\,.
    \label{eq_key_approx}
\end{equation}
It is convenient at this point to define
\begin{equation}
  \tau_{\mathrm{av}}\equiv\dfrac{\ell^2}{2\,D_{\mathrm{av}}},
\end{equation}
which represents the characteristic 
time for a particle with an average diffusion coefficient 
$D_{\mathrm{av}}$ to travel over the distance $\ell$.
As outlined in the Supplemental Material (SM),
taking $p_D^*(D)=\delta(D-D_{\mathrm{av}})$ in
Eq.~\eqref{eq_key_approx} yields
the standard Gaussian result~\cite{weiss1983,oshanin2011,redner2015, lambertW}  
\begin{equation}
\mathbb{E}[T_M]/\tau_{\mathrm{av}}\;\substack{\simeq\\M\gg1}\;1/(2\,\ln
M)
\qquad\textrm{(Gaussian),}
\label{eq_EMFPT_Gauss}
\end{equation}
which highlights how a large number of searchers $M$ reduces the
extreme-MFPT with respect to the typical time taken by a particle to
diffuse over the distance $\ell$.

Let us now focus on classes of subordination processes
$X(t)$ displaying BnG diffusion. These can be introduced via the
stochastic differential equation
\begin{equation}
  \mathrm{d}X(t)
  =\sqrt{2\,D(t)}\,\mathrm{d}B(\mathrm{d}t)\,,
  \label{eq_subordination}
\end{equation}
where $B(t)$ is a Wiener process (Brownian motion) and $D(t)$
describes the fluctuations in time of the diffusion coefficient.  By
defining the \emph{subordinator} as
$S(t)\equiv2\int_0^t\mathrm{d}t'\,D(t')$, Eq.~\eqref{eq_subordination}
can be reformulated in the random path or subordination
representation \mbox{$\mathrm{d}X(t)
=\mathrm{d}B(\mathrm{d}S)$}~\cite{chechkin2017,nampoothiri2021,
nampoothiri2022,marcone2022}.
Depending on the statistical properties of $D(t)$, and hence of the 
subordinator $S(t)$, several stochastic processes can be described by
Eq.~\eqref{eq_subordination}.
For example, if $D(t)=\boldsymbol{Y}^2(t)$ and $\boldsymbol{Y}(t)$
is a $d_{\boldsymbol{Y}}$--dimensional
Ornstein-Uhlenbeck process ($d_{\boldsymbol{Y}}=1,2,3,\dots$), 
we have the DD model~\cite{chechkin2017}.
In the context of  financial markets, under the name of stochastic volatility
models they are used to correct the
Black-Scholes theory for non-Gaussian
effects~\cite{heston1993,sircar2000}.
Another possibility~\cite{nampoothiri2021,nampoothiri2022,marcone2022} is 
$D(t)=D_1/N^\alpha(t)$,
with $\alpha >0 $ and $N(t)\geq1$ a birth-death
process ($N$ only changes by $\pm1$)~\cite{gillespie1992}.
In this case
Eq.~\eqref{eq_subordination} describes the motion of the center of
mass of polymers of size $N$
exchanging monomers with a
chemostat~\cite{deGennes1979,vanderzande1998},
namely
grand canonical polymers (GCP),
and $D_1$ is the diffusion coefficient of a single monomer in solution.
Taking $\alpha=1$ one has the Rouse approximation,
whereas for
the Zimm model $\alpha = \nu$~\cite{Doi1992}, $\nu$ being equal to $1/2$ for ideal,
and $0.588\ldots$ for self-avoiding chains~\cite{deGennes1979}.
This model introduces the concept of critical fluctuations in the
diffusion coefficients, inherited by those of the polymer size $N(t)$
close to the critical point~\cite{deGennes1979,vanderzande1998}.

Distinctive properties of the stochastic process $D(t)$ are its
stationary distribution $p_D^*$,
and the autocorrelation time $\tau$. For the DD model
$\tau$ is a free parameter; in the GCP model
$\tau$ is determined by the reaction rate constants of the birth-death
process and it diverges at criticality
(critical slowing down). Consider a situation in which the
diffusion coefficients of the heterogeneous particles are initially
distributed according to $p_D^*$.
For time $t\ll\tau$ each diffusing particle
retains its initial diffusion coefficient, and the
behavior of the system is described by a statistical average over
$p_D^*$. In the literature, such a superposition of statistics has
been named super-statistics
(SS)~\cite{beck2003,beck2006,hapca2008,wang2012}. During the SS
regime, $p_X(x,t|x_0)$ presents non-Gaussian features like those
reported in Fig.~\ref{fig_excess_pdf}.
On the other hand, as $t\gg\tau$, the probability of the scaled
subordinator $S(t)/t$ concentrates around its average value
$2D_{\mathrm{av}}$, the
central part of $p_X(x,t|x_0)$ becomes Gaussian, and as time passes by
non-Gaussianity is relegated to lesser and lesser probable
fluctuations. This regime is thus associated with a large deviation (LD)
principle~\cite{touchette2009}, and the extreme-MFPT tends to the
behavior in Eq.~\eqref{eq_EMFPT_Gauss}.
The comparison between $\tau$ and $\mathbb{E}[T_M]$ determines whether
the extreme search involves or not BnG features:
The analysis of the extreme-MFPT within the SS (LD) approximation is
applicable to situations where 
$\mathbb{E}[T_M]\ll\tau$ ($\mathbb{E}[T_M]\gg\tau$).

\begin{figure}[t]
  \includegraphics[width=1.0\columnwidth]{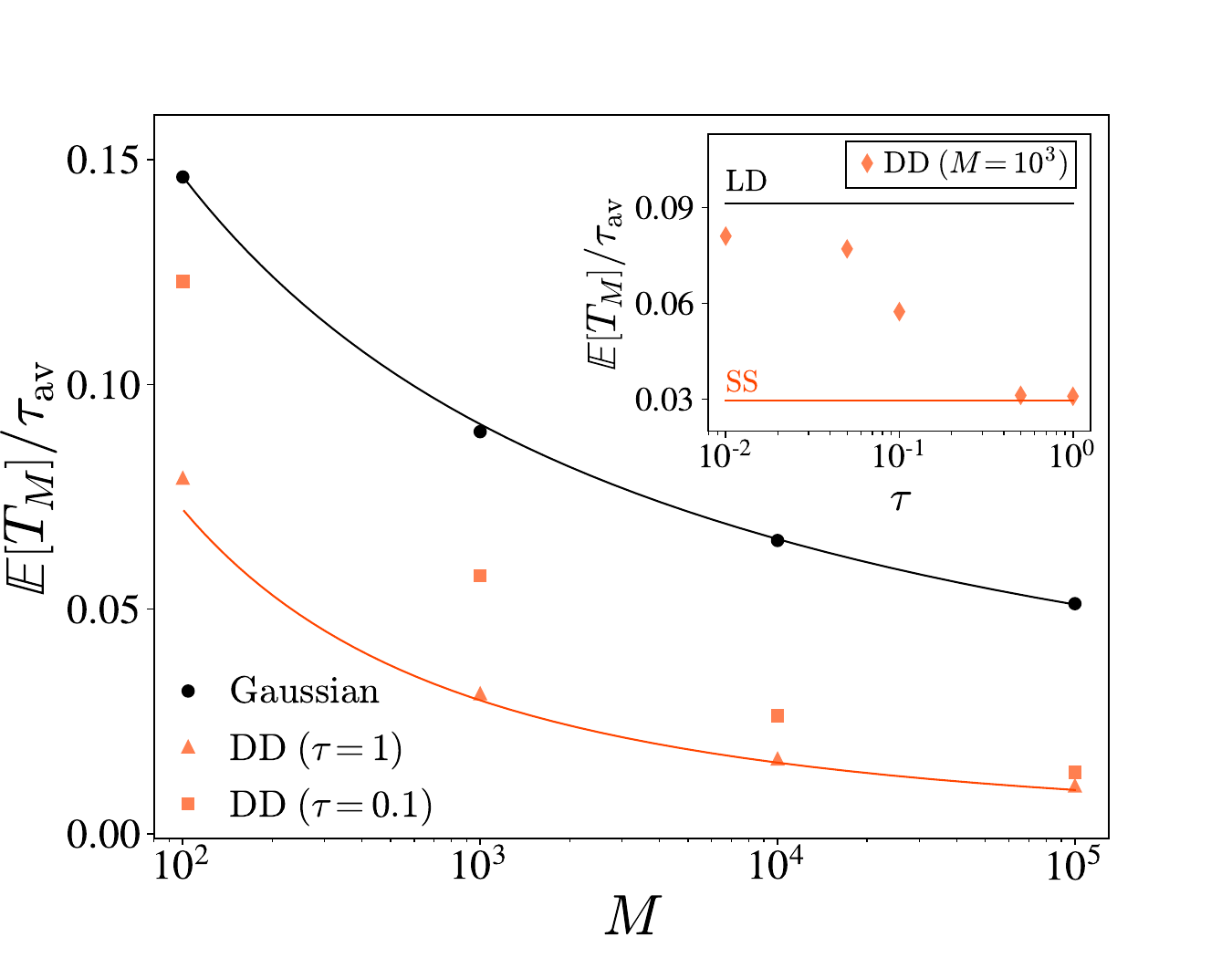}
  \caption{
    Extreme-MFPT for the DD model (orange or gray in grayscale version) and the Gaussian one (black) vs $M$.  
    Note that the ratio $\mathbb{E}[T_M]/\tau_{\mathrm{av}}$ does not depend on $\ell$ and $D_{\mathrm{av}}$.
    Symbols refer to numerical simulations, while solid lines are theoretical estimations from Eqs.~(S8) (black) and~(S16) (orange or gray in grayscale version) of the SM .
    Simulations are with $d_{\boldsymbol{Y}}=1$ and for different values of $\tau$ (details are reported in the SM). 
    Inset: Crossover between the SS and the LD regime, attained upon changing $\tau$ at fixed $M$.}
  \label{fig_extreme_DD}
\end{figure}

In the SM it is reported the steady-state PDF 
$p_D^*(D)$ for the DD model in arbitrary dimension
$d_{\boldsymbol{Y}}$.
Within the SS approximation, we show that 
\begin{equation}
  \mathbb{E}[T_M]/\tau_{\mathrm{av}}\;\substack{\simeq\\M\gg1}
  \; d_{\boldsymbol{Y}} / (\ln M)^2
  \qquad\textrm{(DD model).}
  \label{eq_EMFPT_DD}
\end{equation}
Note that the $1/\ln(M)$ dependence of Eq.~\eqref{eq_EMFPT_Gauss} is
here replaced  by $1/(\ln(M))^2$.
On the contrary, if \mbox{$\mathbb{E}[T_M]\gg\tau$}
Eq.~\eqref{eq_EMFPT_Gauss} applies. 
We thus appreciate that the extra probability for rare large
fluctuations associated with the non-Gaussian tails of the DD model
in the SS regime
drastically reduces the extreme MFPT with respect to Gaussian searches
performed with the average diffusion coefficient.
Such a reduction is particularly visible in Fig.~\ref{fig_extreme_DD} for $\tau=1$, where the 
relation $\mathbb{E}[T_M]\ll\tau$ is satisfied for the whole range of $M$ 
and simulations of the DD model are nicely in agreement with the full
theoretical estimate for the SS regime reported in Eq.~(S16) of the SM.
When $\tau=0.1$, as $M$ decreases a crossover occurs from the SS regime $\sim(\ln M)^2$ to the LD one $\sim\ln M$. 
This is highlighted also in the inset of Fig.~\ref{fig_extreme_DD},
by keeping $M$ fixed and varying $\tau$.

\begin{figure}[t]
  \includegraphics[width=0.9\columnwidth]{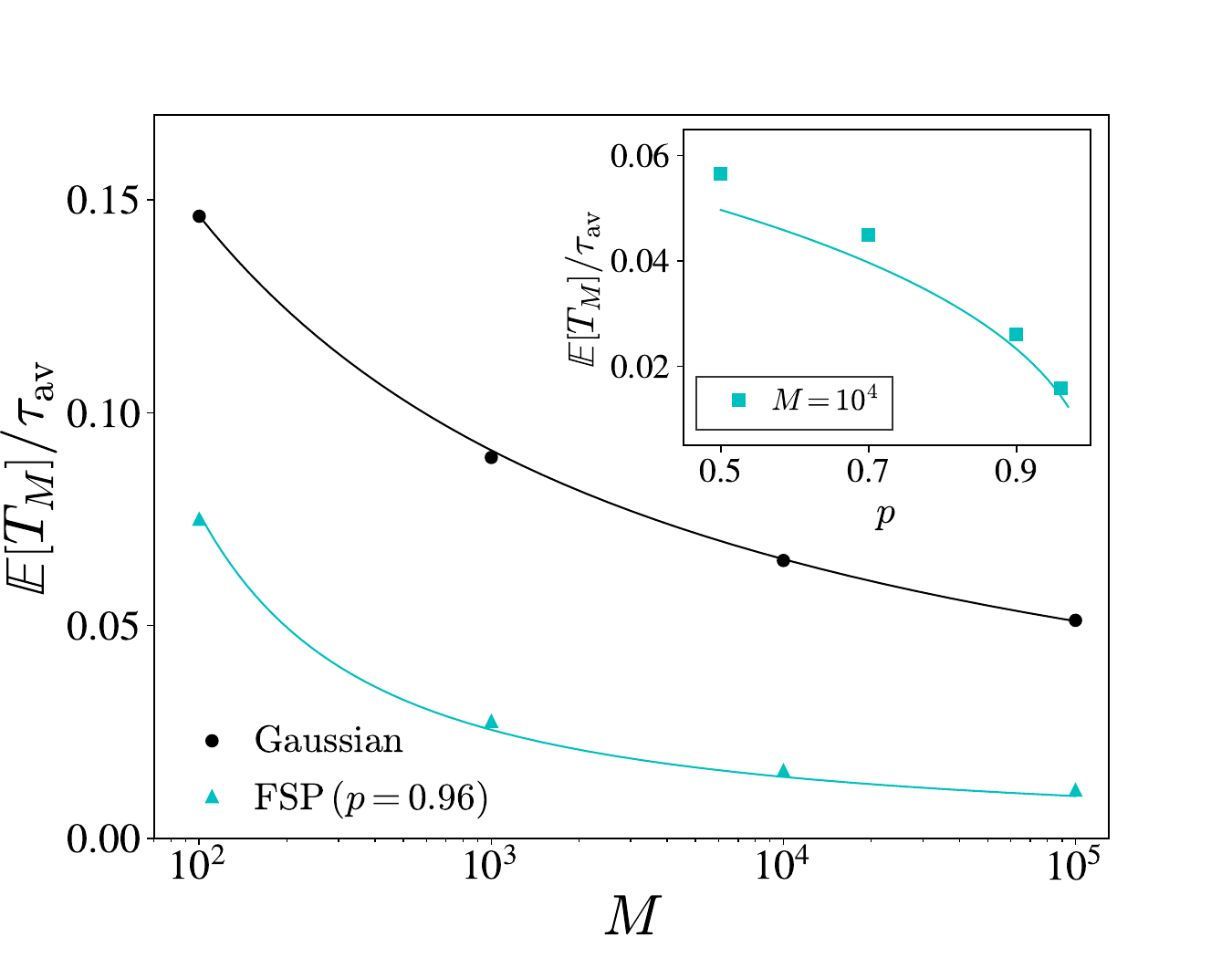}
  \caption{
  Extreme-MFPT for FSP model (cyan or gray in grayscale version). In analogy to Fig.~\ref{fig_extreme_DD}, 
  symbols refer to numerical simulations, while solid lines are theoretical 
  estimations from Eqs.~(S8) (black) and~(S23) (cyan or gray in grayscale version) of the SM.
  Inset: behavior with respect to $p$ at fixed $M$; the solid line 
  indicates the theoretical trend from Eq.~\eqref{eq_EMFPT_FSP}.
  Simulation details are reported in the SM.
  }
  \label{fig_extreme_FP}
\end{figure}

The SS regime for the GCP model has a simple, practical experimental
implementation: A polydisperse sample produced 
in a step-growth polymerization~\cite{odian2004}.
Whereas for GCPs the
polymerization/depolymerization process continuously occurs over time while 
system and chemostat exchange monomers, in the polydisperse case
polymerization terminates after the initial outgrowth and the sample 
constantly remains in the SS regime with $D$ a static random variable.
Taking for simplicity chains with exactly one 
reaction center in the end, one can
equivalently address the SS regime of GCPs 
considering a heterogeneous sample
distributed according to the Flory-Schulz 
size distribution~\cite{flory1953} $p_N^*(n)=(1-p)\,p^{n-1}$, where 
$0\leq p\leq1$ is the polymerization extent. We will refer to this 
as Flory-Schulz polydisperse (FSP) model.
As $p\to 1^-$, the average polymer size
$\mathbb{E}[N]=1/(1-p)$ diverges and the system becomes
critical~\cite{deGennes1979,vanderzande1998,nampoothiri2022}.
Correspondingly, $D_{\mathrm{av}}$ tends to zero and
$\tau_{\mathrm{av}}$ diverges. 
The analysis reported in the SM for Rouse polymers yields
\begin{equation}
  \dfrac{\mathbb{E}[T_M]}{\tau_{\mathrm{av}}}
  \;\substack{\simeq\\M\,(1-p)\gg1}
  \;-\dfrac{(1-p)\,\ln(1-p)}{2\,p\,\ln((1-p)\,M)}
  \quad\textrm{(FSP model),}
  \label{eq_EMFPT_FSP}
\end{equation}
where, in consistency with our approximations, we have assumed
a sufficiently large
number of searchers such that $M\gg(1-p)^{-1}$.
In the LD regime, the extreme MFPT of GCPs is again
described by 
Eq.~\eqref{eq_EMFPT_Gauss}.
Comparison of Eq.~\eqref{eq_EMFPT_FSP} with 
Eq.~\eqref{eq_EMFPT_Gauss} reveals that while Gaussian searches take
an infinite time to be accomplished as the system approaches
criticality and $\tau_{\mathrm{av}}$ diverges, non-Gaussian ones are
still realized within a finite time.
This means that wild fluctuations in the polymer size induce such heavy 
tail effect in $p_X(x,t|x_0)$ to keep the extreme-MFPT finite, eluding the
critical slowing down for this kind of search.
One might argue that since this analysis applies to the center of
mass of the polymer which is an immaterial point in space, is of
limited practical relevance. However, our results indicate that the
instances which reach the target under non-Gaussian
heterogeneous conditions are precisely those fast diffusers
responsible for the ``tail effect''. These are the polymers with a
small size, for which the Rouse time of the chain~\cite{Doi1992} is
very small, and hence their center-of-mass dynamical time-scale
corresponds to that of any monomer unit acting in practice
as ligand. 
We may add that this example shows that heterogeneity supplements
to extreme-searches the concept of \textit{fitness}:
In a heterogeneous sample not only the geodesic path to the target is
followed in extreme searches, but the successful searcher happens to
belong to the fittest subset (in our case, the fastest, small-size
polymers). 
In analogy with the previous plots, simulations in Fig.~\ref{fig_extreme_FP} 
confirm our analytical predictions for the FSP model.

It is interesting to point out that, at variance with the DD model,
the extreme-MFPT for the polydisperse polymers
displays the same $1/\ln(M)$
dependence of the Gaussian case.  This is to be ascribed to the sharp
large-value cutoff at $D_n=D_1$ of the $p^*_D(D_n)$ distribution.  To
clarify this point, we have analyzed a class of generalized
gamma distributions~\cite{sposini2018} $p^*_D(D)
\substack{\sim\\D\to\infty} \exp[-(A(\nu, \eta)
  D/D_\mathrm{av})^\eta]$, with the parameter $\eta$ characterizing
different tail behaviors.  The same procedure used for the other models gives
\begin{equation}
    \mathbb{E}[T_M]/\tau_{\mathrm{av}}\;\substack{\simeq\\M\gg1}\; 
    \left[D(\nu,\eta)/\ln(M)\right]^{\frac{(\eta+1)}{\eta}}
    \, \textrm{(gen Gamma model),}
    \label{eq_EMFPT_genDD}
\end{equation} 
where the full expression of the coefficient $D(\eta,\nu)$ is provided in 
the SM. On the one hand, for $\eta =1 $ (exponential tail) we recover the 
DD result, namely $1/(\ln(M))^2$. On the other hand, in the limit 
$\eta\to\infty$ for which the tail of $p^*_D(D)$ drops sharply approaching 
the step function decay of the FSP model,
we recover the $1/\ln(M)$ dependence. The result in Eq.~\eqref{eq_EMFPT_genDD} 
shows explicitly that: a) There is no universal behavior in $M$ when we move 
outside of the Gaussian regime; b) For random diffusivity model it is 
the tail of the diffusivity distribution that decides such a trend.  

The origin of non-Gaussianity that we have addressed is amenable to
the heterogeneity of the ensemble of diffusers and/or of the
environment~\cite{sokolov2021}. Such heterogeneity implies both an
excess of probability in the central part and in the tails of the
displacements distribution, when compared with the Gaussian
one~\cite{sposini2023pre}.
We have shown that a higher probability for few, faster diffusers
(``tail effect'') influences extreme searches, pointing out that a
redundant information stored in \textit{diverse searchers} strongly
enhances the fast targeting of the first few instances.
Non-Gaussianity is both disadvantageous~\cite{sposini2023pre}
and advantageous. It is disadvantageous when the activation of a
biological function needs a large percentage of ligands to bind
receptors; it is advantageous when only a few searchers, among many,
are required to reach the target. The latter is the typical situation
in which diffusing particles are carriers of information, like in
human reproduction.  For this kind of search, diversity appears to
be an efficient strategy to be recognized in evolutionary examples,
and exploited in the design of efficient deliveries.  A
straightforward setup for experimental confirmation of our results is
that of polydisperse polymers.

More broadly, we expect this investigation to open prospects in
understanding the role of heterogeneity in diffusion transport
phenomena, for instance in models where the BnG behavior has been
discovered~\cite{wang2012,toyota2011,yu2013,yu2014,chakraborty2020,
  weeks2000,wagner2017,jeon2016,yamamoto2017,stylianidou2014,parry2014,munder2016,
  cherstvy2018,li2019,cuetos2018,hapca2008,pastore2021rapid,giavazzi2022},
and likewise in the world of anomalous processes where the mean
squared displacement grows non-linearly in time. Indeed, recent
single-particle tracking experiments in crowded environments -- such as
those of biological cells -- show that heterogeneity manifests itself
not only in the variability of transport
coefficients~\cite{wang2020unexpected}, but also in fluctuations of
the anomalous diffusion
exponent~\cite{sadoon2018anomalous,cherstvy2019non,
  sabri2020elucidating, han2020deciphering, benelli2021sub,
  janczura2021identifying, korabel2021local,
  speckner2021single,balcerek2022fractional,wang2023memory,
  korabel2023ensemble}.

\section*{Acknowledgments}
F.S. and F.B. acknowledge the support by the project MUR-PRIN 2022ETXBEY, "Fickian non-Gaussian diffusion in static and dynamic
environments", funded by the European Union – Next Generation EU.
V.S. acknowledges the support from the 
European Commission through the Marie Sk\l{}odowska-Curie COFUND 
project REWIRE, grant agreement No. 847693. A.C. acknowledges the support of the Polish National Agency for Academic Exchange (NAWA).


\begin{thebibliography}{75}%
\makeatletter
\providecommand \@ifxundefined [1]{%
 \@ifx{#1\undefined}
}%
\providecommand \@ifnum [1]{%
 \ifnum #1\expandafter \@firstoftwo
 \else \expandafter \@secondoftwo
 \fi
}%
\providecommand \@ifx [1]{%
 \ifx #1\expandafter \@firstoftwo
 \else \expandafter \@secondoftwo
 \fi
}%
\providecommand \natexlab [1]{#1}%
\providecommand \enquote  [1]{``#1''}%
\providecommand \bibnamefont  [1]{#1}%
\providecommand \bibfnamefont [1]{#1}%
\providecommand \citenamefont [1]{#1}%
\providecommand \href@noop [0]{\@secondoftwo}%
\providecommand \href [0]{\begingroup \@sanitize@url \@href}%
\providecommand \@href[1]{\@@startlink{#1}\@@href}%
\providecommand \@@href[1]{\endgroup#1\@@endlink}%
\providecommand \@sanitize@url [0]{\catcode `\\12\catcode `\$12\catcode
  `\&12\catcode `\#12\catcode `\^12\catcode `\_12\catcode `\%12\relax}%
\providecommand \@@startlink[1]{}%
\providecommand \@@endlink[0]{}%
\providecommand \url  [0]{\begingroup\@sanitize@url \@url }%
\providecommand \@url [1]{\endgroup\@href {#1}{\urlprefix }}%
\providecommand \urlprefix  [0]{URL }%
\providecommand \Eprint [0]{\href }%
\providecommand \doibase [0]{http://dx.doi.org/}%
\providecommand \selectlanguage [0]{\@gobble}%
\providecommand \bibinfo  [0]{\@secondoftwo}%
\providecommand \bibfield  [0]{\@secondoftwo}%
\providecommand \translation [1]{[#1]}%
\providecommand \BibitemOpen [0]{}%
\providecommand \bibitemStop [0]{}%
\providecommand \bibitemNoStop [0]{.\EOS\space}%
\providecommand \EOS [0]{\spacefactor3000\relax}%
\providecommand \BibitemShut  [1]{\csname bibitem#1\endcsname}%
\let\auto@bib@innerbib\@empty
\bibitem [{\citenamefont {Wang}\ \emph {et~al.}(2009)\citenamefont {Wang},
  \citenamefont {Anthony}, \citenamefont {Bae},\ and\ \citenamefont
  {Granick}}]{wang2009}%
  \BibitemOpen
  \bibfield  {author} {\bibinfo {author} {\bibfnamefont {B.}~\bibnamefont
  {Wang}}, \bibinfo {author} {\bibfnamefont {S.~M.}\ \bibnamefont {Anthony}},
  \bibinfo {author} {\bibfnamefont {S.~C.}\ \bibnamefont {Bae}}, \ and\
  \bibinfo {author} {\bibfnamefont {S.}~\bibnamefont {Granick}},\ }\href@noop
  {} {\bibfield  {journal} {\bibinfo  {journal} {Proceedings of the National
  Academy of Sciences}\ }\textbf {\bibinfo {volume} {106}},\ \bibinfo {pages}
  {15160} (\bibinfo {year} {2009})}\BibitemShut {NoStop}%
\bibitem [{\citenamefont {Wang}\ \emph {et~al.}(2012)\citenamefont {Wang},
  \citenamefont {Kuo}, \citenamefont {Bae},\ and\ \citenamefont
  {Granick}}]{wang2012}%
  \BibitemOpen
  \bibfield  {author} {\bibinfo {author} {\bibfnamefont {B.}~\bibnamefont
  {Wang}}, \bibinfo {author} {\bibfnamefont {J.}~\bibnamefont {Kuo}}, \bibinfo
  {author} {\bibfnamefont {S.~C.}\ \bibnamefont {Bae}}, \ and\ \bibinfo
  {author} {\bibfnamefont {S.}~\bibnamefont {Granick}},\ }\href@noop {}
  {\bibfield  {journal} {\bibinfo  {journal} {Nature materials}\ }\textbf
  {\bibinfo {volume} {11}},\ \bibinfo {pages} {481} (\bibinfo {year}
  {2012})}\BibitemShut {NoStop}%
\bibitem [{\citenamefont {Toyota}\ \emph {et~al.}(2011)\citenamefont {Toyota},
  \citenamefont {Head}, \citenamefont {Schmidt},\ and\ \citenamefont
  {Mizuno}}]{toyota2011}%
  \BibitemOpen
  \bibfield  {author} {\bibinfo {author} {\bibfnamefont {T.}~\bibnamefont
  {Toyota}}, \bibinfo {author} {\bibfnamefont {D.~A.}\ \bibnamefont {Head}},
  \bibinfo {author} {\bibfnamefont {C.~F.}\ \bibnamefont {Schmidt}}, \ and\
  \bibinfo {author} {\bibfnamefont {D.}~\bibnamefont {Mizuno}},\ }\href@noop {}
  {\bibfield  {journal} {\bibinfo  {journal} {Soft Matter}\ }\textbf {\bibinfo
  {volume} {7}},\ \bibinfo {pages} {3234} (\bibinfo {year} {2011})}\BibitemShut
  {NoStop}%
\bibitem [{\citenamefont {Yu}\ \emph {et~al.}(2013)\citenamefont {Yu},
  \citenamefont {Guan}, \citenamefont {Chen}, \citenamefont {Bae},\ and\
  \citenamefont {Granick}}]{yu2013}%
  \BibitemOpen
  \bibfield  {author} {\bibinfo {author} {\bibfnamefont {C.}~\bibnamefont
  {Yu}}, \bibinfo {author} {\bibfnamefont {J.}~\bibnamefont {Guan}}, \bibinfo
  {author} {\bibfnamefont {K.}~\bibnamefont {Chen}}, \bibinfo {author}
  {\bibfnamefont {S.~C.}\ \bibnamefont {Bae}}, \ and\ \bibinfo {author}
  {\bibfnamefont {S.}~\bibnamefont {Granick}},\ }\href@noop {} {\bibfield
  {journal} {\bibinfo  {journal} {ACS Nano}\ }\textbf {\bibinfo {volume} {7}},\
  \bibinfo {pages} {9735} (\bibinfo {year} {2013})}\BibitemShut {NoStop}%
\bibitem [{\citenamefont {Yu}\ and\ \citenamefont {Granick}(2014)}]{yu2014}%
  \BibitemOpen
  \bibfield  {author} {\bibinfo {author} {\bibfnamefont {C.}~\bibnamefont
  {Yu}}\ and\ \bibinfo {author} {\bibfnamefont {S.}~\bibnamefont {Granick}},\
  }\href@noop {} {\bibfield  {journal} {\bibinfo  {journal} {Langmuir}\
  }\textbf {\bibinfo {volume} {30}},\ \bibinfo {pages} {14538} (\bibinfo {year}
  {2014})}\BibitemShut {NoStop}%
\bibitem [{\citenamefont {Chakraborty}\ and\ \citenamefont
  {Roichman}(2020)}]{chakraborty2020}%
  \BibitemOpen
  \bibfield  {author} {\bibinfo {author} {\bibfnamefont {I.}~\bibnamefont
  {Chakraborty}}\ and\ \bibinfo {author} {\bibfnamefont {Y.}~\bibnamefont
  {Roichman}},\ }\href@noop {} {\bibfield  {journal} {\bibinfo  {journal}
  {Physical Review Research}\ }\textbf {\bibinfo {volume} {2}},\ \bibinfo
  {pages} {022020} (\bibinfo {year} {2020})}\BibitemShut {NoStop}%
\bibitem [{\citenamefont {Weeks}\ \emph {et~al.}(2000)\citenamefont {Weeks},
  \citenamefont {Crocker}, \citenamefont {Levitt}, \citenamefont {Schofield},\
  and\ \citenamefont {Weitz}}]{weeks2000}%
  \BibitemOpen
  \bibfield  {author} {\bibinfo {author} {\bibfnamefont {E.~R.}\ \bibnamefont
  {Weeks}}, \bibinfo {author} {\bibfnamefont {J.~C.}\ \bibnamefont {Crocker}},
  \bibinfo {author} {\bibfnamefont {A.~C.}\ \bibnamefont {Levitt}}, \bibinfo
  {author} {\bibfnamefont {A.}~\bibnamefont {Schofield}}, \ and\ \bibinfo
  {author} {\bibfnamefont {D.~A.}\ \bibnamefont {Weitz}},\ }\href@noop {}
  {\bibfield  {journal} {\bibinfo  {journal} {Science}\ }\textbf {\bibinfo
  {volume} {287}},\ \bibinfo {pages} {627} (\bibinfo {year}
  {2000})}\BibitemShut {NoStop}%
\bibitem [{\citenamefont {Wagner}\ \emph {et~al.}(2017)\citenamefont {Wagner},
  \citenamefont {Turner}, \citenamefont {Rubinstein}, \citenamefont
  {McKinley},\ and\ \citenamefont {Ribbeck}}]{wagner2017}%
  \BibitemOpen
  \bibfield  {author} {\bibinfo {author} {\bibfnamefont {C.~E.}\ \bibnamefont
  {Wagner}}, \bibinfo {author} {\bibfnamefont {B.~S.}\ \bibnamefont {Turner}},
  \bibinfo {author} {\bibfnamefont {M.}~\bibnamefont {Rubinstein}}, \bibinfo
  {author} {\bibfnamefont {G.~H.}\ \bibnamefont {McKinley}}, \ and\ \bibinfo
  {author} {\bibfnamefont {K.}~\bibnamefont {Ribbeck}},\ }\href@noop {}
  {\bibfield  {journal} {\bibinfo  {journal} {Biomacromolecules}\ }\textbf
  {\bibinfo {volume} {18}},\ \bibinfo {pages} {3654} (\bibinfo {year}
  {2017})}\BibitemShut {NoStop}%
\bibitem [{\citenamefont {Jeon}\ \emph {et~al.}(2016)\citenamefont {Jeon},
  \citenamefont {Javanainen}, \citenamefont {Martinez-Seara}, \citenamefont
  {Metzler},\ and\ \citenamefont {Vattulainen}}]{jeon2016}%
  \BibitemOpen
  \bibfield  {author} {\bibinfo {author} {\bibfnamefont {J.-H.}\ \bibnamefont
  {Jeon}}, \bibinfo {author} {\bibfnamefont {M.}~\bibnamefont {Javanainen}},
  \bibinfo {author} {\bibfnamefont {H.}~\bibnamefont {Martinez-Seara}},
  \bibinfo {author} {\bibfnamefont {R.}~\bibnamefont {Metzler}}, \ and\
  \bibinfo {author} {\bibfnamefont {I.}~\bibnamefont {Vattulainen}},\
  }\href@noop {} {\bibfield  {journal} {\bibinfo  {journal} {Physical Review
  X}\ }\textbf {\bibinfo {volume} {6}},\ \bibinfo {pages} {021006} (\bibinfo
  {year} {2016})}\BibitemShut {NoStop}%
\bibitem [{\citenamefont {Yamamoto}\ \emph {et~al.}(2017)\citenamefont
  {Yamamoto}, \citenamefont {Akimoto}, \citenamefont {Kalli}, \citenamefont
  {Yasuoka},\ and\ \citenamefont {Sansom}}]{yamamoto2017}%
  \BibitemOpen
  \bibfield  {author} {\bibinfo {author} {\bibfnamefont {E.}~\bibnamefont
  {Yamamoto}}, \bibinfo {author} {\bibfnamefont {T.}~\bibnamefont {Akimoto}},
  \bibinfo {author} {\bibfnamefont {A.~C.}\ \bibnamefont {Kalli}}, \bibinfo
  {author} {\bibfnamefont {K.}~\bibnamefont {Yasuoka}}, \ and\ \bibinfo
  {author} {\bibfnamefont {M.~S.}\ \bibnamefont {Sansom}},\ }\href@noop {}
  {\bibfield  {journal} {\bibinfo  {journal} {Science advances}\ }\textbf
  {\bibinfo {volume} {3}},\ \bibinfo {pages} {e1601871} (\bibinfo {year}
  {2017})}\BibitemShut {NoStop}%
\bibitem [{\citenamefont {Stylianidou}\ \emph {et~al.}(2014)\citenamefont
  {Stylianidou}, \citenamefont {Kuwada},\ and\ \citenamefont
  {Wiggins}}]{stylianidou2014}%
  \BibitemOpen
  \bibfield  {author} {\bibinfo {author} {\bibfnamefont {S.}~\bibnamefont
  {Stylianidou}}, \bibinfo {author} {\bibfnamefont {N.~J.}\ \bibnamefont
  {Kuwada}}, \ and\ \bibinfo {author} {\bibfnamefont {P.~A.}\ \bibnamefont
  {Wiggins}},\ }\href@noop {} {\bibfield  {journal} {\bibinfo  {journal}
  {Biophysical journal}\ }\textbf {\bibinfo {volume} {107}},\ \bibinfo {pages}
  {2684} (\bibinfo {year} {2014})}\BibitemShut {NoStop}%
\bibitem [{\citenamefont {Parry}\ \emph {et~al.}(2014)\citenamefont {Parry},
  \citenamefont {Surovtsev}, \citenamefont {Cabeen}, \citenamefont {O’Hern},
  \citenamefont {Dufresne},\ and\ \citenamefont {Jacobs-Wagner}}]{parry2014}%
  \BibitemOpen
  \bibfield  {author} {\bibinfo {author} {\bibfnamefont {B.~R.}\ \bibnamefont
  {Parry}}, \bibinfo {author} {\bibfnamefont {I.~V.}\ \bibnamefont
  {Surovtsev}}, \bibinfo {author} {\bibfnamefont {M.~T.}\ \bibnamefont
  {Cabeen}}, \bibinfo {author} {\bibfnamefont {C.~S.}\ \bibnamefont
  {O’Hern}}, \bibinfo {author} {\bibfnamefont {E.~R.}\ \bibnamefont
  {Dufresne}}, \ and\ \bibinfo {author} {\bibfnamefont {C.}~\bibnamefont
  {Jacobs-Wagner}},\ }\href@noop {} {\bibfield  {journal} {\bibinfo  {journal}
  {Cell}\ }\textbf {\bibinfo {volume} {156}},\ \bibinfo {pages} {183} (\bibinfo
  {year} {2014})}\BibitemShut {NoStop}%
\bibitem [{\citenamefont {Munder}\ \emph {et~al.}(2016)\citenamefont {Munder},
  \citenamefont {Midtvedt}, \citenamefont {Franzmann}, \citenamefont {Nuske},
  \citenamefont {Otto}, \citenamefont {Herbig}, \citenamefont {Ulbricht},
  \citenamefont {M{\"u}ller}, \citenamefont {Taubenberger}, \citenamefont
  {Maharana} \emph {et~al.}}]{munder2016}%
  \BibitemOpen
  \bibfield  {author} {\bibinfo {author} {\bibfnamefont {M.~C.}\ \bibnamefont
  {Munder}}, \bibinfo {author} {\bibfnamefont {D.}~\bibnamefont {Midtvedt}},
  \bibinfo {author} {\bibfnamefont {T.}~\bibnamefont {Franzmann}}, \bibinfo
  {author} {\bibfnamefont {E.}~\bibnamefont {Nuske}}, \bibinfo {author}
  {\bibfnamefont {O.}~\bibnamefont {Otto}}, \bibinfo {author} {\bibfnamefont
  {M.}~\bibnamefont {Herbig}}, \bibinfo {author} {\bibfnamefont
  {E.}~\bibnamefont {Ulbricht}}, \bibinfo {author} {\bibfnamefont
  {P.}~\bibnamefont {M{\"u}ller}}, \bibinfo {author} {\bibfnamefont
  {A.}~\bibnamefont {Taubenberger}}, \bibinfo {author} {\bibfnamefont
  {S.}~\bibnamefont {Maharana}},  \emph {et~al.},\ }\href@noop {} {\bibfield
  {journal} {\bibinfo  {journal} {elife}\ }\textbf {\bibinfo {volume} {5}},\
  \bibinfo {pages} {e09347} (\bibinfo {year} {2016})}\BibitemShut {NoStop}%
\bibitem [{\citenamefont {Cherstvy}\ \emph {et~al.}(2018)\citenamefont
  {Cherstvy}, \citenamefont {Nagel}, \citenamefont {Beta},\ and\ \citenamefont
  {Metzler}}]{cherstvy2018}%
  \BibitemOpen
  \bibfield  {author} {\bibinfo {author} {\bibfnamefont {A.~G.}\ \bibnamefont
  {Cherstvy}}, \bibinfo {author} {\bibfnamefont {O.}~\bibnamefont {Nagel}},
  \bibinfo {author} {\bibfnamefont {C.}~\bibnamefont {Beta}}, \ and\ \bibinfo
  {author} {\bibfnamefont {R.}~\bibnamefont {Metzler}},\ }\href@noop {}
  {\bibfield  {journal} {\bibinfo  {journal} {Physical Chemistry Chemical
  Physics}\ }\textbf {\bibinfo {volume} {20}},\ \bibinfo {pages} {23034}
  (\bibinfo {year} {2018})}\BibitemShut {NoStop}%
\bibitem [{\citenamefont {Li}\ \emph {et~al.}(2019)\citenamefont {Li},
  \citenamefont {Marchesoni}, \citenamefont {Debnath},\ and\ \citenamefont
  {Ghosh}}]{li2019}%
  \BibitemOpen
  \bibfield  {author} {\bibinfo {author} {\bibfnamefont {Y.}~\bibnamefont
  {Li}}, \bibinfo {author} {\bibfnamefont {F.}~\bibnamefont {Marchesoni}},
  \bibinfo {author} {\bibfnamefont {D.}~\bibnamefont {Debnath}}, \ and\
  \bibinfo {author} {\bibfnamefont {P.~K.}\ \bibnamefont {Ghosh}},\ }\href@noop
  {} {\bibfield  {journal} {\bibinfo  {journal} {Physical Review Research}\
  }\textbf {\bibinfo {volume} {1}},\ \bibinfo {pages} {033003} (\bibinfo {year}
  {2019})}\BibitemShut {NoStop}%
\bibitem [{\citenamefont {Cuetos}\ \emph {et~al.}(2018)\citenamefont {Cuetos},
  \citenamefont {Morillo},\ and\ \citenamefont {Patti}}]{cuetos2018}%
  \BibitemOpen
  \bibfield  {author} {\bibinfo {author} {\bibfnamefont {A.}~\bibnamefont
  {Cuetos}}, \bibinfo {author} {\bibfnamefont {N.}~\bibnamefont {Morillo}}, \
  and\ \bibinfo {author} {\bibfnamefont {A.}~\bibnamefont {Patti}},\
  }\href@noop {} {\bibfield  {journal} {\bibinfo  {journal} {Physical Review
  E}\ }\textbf {\bibinfo {volume} {98}},\ \bibinfo {pages} {042129} (\bibinfo
  {year} {2018})}\BibitemShut {NoStop}%
\bibitem [{\citenamefont {Hapca}\ \emph {et~al.}(2008)\citenamefont {Hapca},
  \citenamefont {Crawford},\ and\ \citenamefont {Young}}]{hapca2008}%
  \BibitemOpen
  \bibfield  {author} {\bibinfo {author} {\bibfnamefont {S.}~\bibnamefont
  {Hapca}}, \bibinfo {author} {\bibfnamefont {J.~W.}\ \bibnamefont {Crawford}},
  \ and\ \bibinfo {author} {\bibfnamefont {I.~M.}\ \bibnamefont {Young}},\
  }\href@noop {} {\bibfield  {journal} {\bibinfo  {journal} {Journal of the
  Royal Society Interface}\ }\textbf {\bibinfo {volume} {6}},\ \bibinfo {pages}
  {111} (\bibinfo {year} {2008})}\BibitemShut {NoStop}%
\bibitem [{\citenamefont {Pastore}\ \emph {et~al.}(2021)\citenamefont
  {Pastore}, \citenamefont {Ciarlo}, \citenamefont {Pesce}, \citenamefont
  {Greco},\ and\ \citenamefont {Sasso}}]{pastore2021rapid}%
  \BibitemOpen
  \bibfield  {author} {\bibinfo {author} {\bibfnamefont {R.}~\bibnamefont
  {Pastore}}, \bibinfo {author} {\bibfnamefont {A.}~\bibnamefont {Ciarlo}},
  \bibinfo {author} {\bibfnamefont {G.}~\bibnamefont {Pesce}}, \bibinfo
  {author} {\bibfnamefont {F.}~\bibnamefont {Greco}}, \ and\ \bibinfo {author}
  {\bibfnamefont {A.}~\bibnamefont {Sasso}},\ }\href@noop {} {\bibfield
  {journal} {\bibinfo  {journal} {Physical Review Letters}\ }\textbf {\bibinfo
  {volume} {126}},\ \bibinfo {pages} {158003} (\bibinfo {year}
  {2021})}\BibitemShut {NoStop}%
\bibitem [{\citenamefont {Brizioli}\ \emph {et~al.}(2022)\citenamefont
  {Brizioli}, \citenamefont {Sentjabrskaja}, \citenamefont {Egelhaaf},
  \citenamefont {Laurati}, \citenamefont {Cerbino},\ and\ \citenamefont
  {Giavazzi}}]{giavazzi2022}%
  \BibitemOpen
  \bibfield  {author} {\bibinfo {author} {\bibfnamefont {M.}~\bibnamefont
  {Brizioli}}, \bibinfo {author} {\bibfnamefont {T.}~\bibnamefont
  {Sentjabrskaja}}, \bibinfo {author} {\bibfnamefont {S.~U.}\ \bibnamefont
  {Egelhaaf}}, \bibinfo {author} {\bibfnamefont {M.}~\bibnamefont {Laurati}},
  \bibinfo {author} {\bibfnamefont {R.}~\bibnamefont {Cerbino}}, \ and\
  \bibinfo {author} {\bibfnamefont {F.}~\bibnamefont {Giavazzi}},\ }\href@noop
  {} {\bibfield  {journal} {\bibinfo  {journal} {Frontiers in Physics}\
  }\textbf {\bibinfo {volume} {10}},\ \bibinfo {pages} {893777} (\bibinfo
  {year} {2022})}\BibitemShut {NoStop}%
\bibitem [{\citenamefont {Alexandre}\ \emph {et~al.}(2023)\citenamefont
  {Alexandre}, \citenamefont {Lavaud}, \citenamefont {Fares}, \citenamefont
  {Millan}, \citenamefont {Louyer}, \citenamefont {Salez}, \citenamefont
  {Amarouchene}, \citenamefont {Gu\'erin},\ and\ \citenamefont
  {Dean}}]{dean2023}%
  \BibitemOpen
  \bibfield  {author} {\bibinfo {author} {\bibfnamefont {A.}~\bibnamefont
  {Alexandre}}, \bibinfo {author} {\bibfnamefont {M.}~\bibnamefont {Lavaud}},
  \bibinfo {author} {\bibfnamefont {N.}~\bibnamefont {Fares}}, \bibinfo
  {author} {\bibfnamefont {E.}~\bibnamefont {Millan}}, \bibinfo {author}
  {\bibfnamefont {Y.}~\bibnamefont {Louyer}}, \bibinfo {author} {\bibfnamefont
  {T.}~\bibnamefont {Salez}}, \bibinfo {author} {\bibfnamefont
  {Y.}~\bibnamefont {Amarouchene}}, \bibinfo {author} {\bibfnamefont
  {T.}~\bibnamefont {Gu\'erin}}, \ and\ \bibinfo {author} {\bibfnamefont
  {D.~S.}\ \bibnamefont {Dean}},\ }\href@noop {} {\bibfield  {journal}
  {\bibinfo  {journal} {Phys. Rev. Lett.}\ }\textbf {\bibinfo {volume} {130}},\
  \bibinfo {pages} {077101} (\bibinfo {year} {2023})}\BibitemShut {NoStop}%
\bibitem [{\citenamefont {Pastore}\ and\ \citenamefont
  {Raos}(2015)}]{pastore2015}%
  \BibitemOpen
  \bibfield  {author} {\bibinfo {author} {\bibfnamefont {R.}~\bibnamefont
  {Pastore}}\ and\ \bibinfo {author} {\bibfnamefont {G.}~\bibnamefont {Raos}},\
  }\href@noop {} {\bibfield  {journal} {\bibinfo  {journal} {Soft Matter}\
  }\textbf {\bibinfo {volume} {11}},\ \bibinfo {pages} {8083} (\bibinfo {year}
  {2015})}\BibitemShut {NoStop}%
\bibitem [{\citenamefont {Miotto}\ \emph {et~al.}(2021)\citenamefont {Miotto},
  \citenamefont {Pigolotti}, \citenamefont {Chechkin},\ and\ \citenamefont
  {Rold{\'a}n-Vargas}}]{miotto2021length}%
  \BibitemOpen
  \bibfield  {author} {\bibinfo {author} {\bibfnamefont {J.~M.}\ \bibnamefont
  {Miotto}}, \bibinfo {author} {\bibfnamefont {S.}~\bibnamefont {Pigolotti}},
  \bibinfo {author} {\bibfnamefont {A.~V.}\ \bibnamefont {Chechkin}}, \ and\
  \bibinfo {author} {\bibfnamefont {S.}~\bibnamefont {Rold{\'a}n-Vargas}},\
  }\href@noop {} {\bibfield  {journal} {\bibinfo  {journal} {Physical Review
  X}\ }\textbf {\bibinfo {volume} {11}},\ \bibinfo {pages} {031002} (\bibinfo
  {year} {2021})}\BibitemShut {NoStop}%
\bibitem [{\citenamefont {Rusciano}\ \emph {et~al.}(2022)\citenamefont
  {Rusciano}, \citenamefont {Pastore},\ and\ \citenamefont
  {Greco}}]{pastore2022}%
  \BibitemOpen
  \bibfield  {author} {\bibinfo {author} {\bibfnamefont {F.}~\bibnamefont
  {Rusciano}}, \bibinfo {author} {\bibfnamefont {R.}~\bibnamefont {Pastore}}, \
  and\ \bibinfo {author} {\bibfnamefont {F.}~\bibnamefont {Greco}},\
  }\href@noop {} {\bibfield  {journal} {\bibinfo  {journal} {Physical Review
  Letters}\ }\textbf {\bibinfo {volume} {128}},\ \bibinfo {pages} {168001}
  (\bibinfo {year} {2022})}\BibitemShut {NoStop}%
\bibitem [{\citenamefont {Lanoisel\'ee}\ \emph {et~al.}(2018)\citenamefont
  {Lanoisel\'ee}, \citenamefont {Moutal},\ and\ \citenamefont
  {Grebenkov}}]{grebenkov2018}%
  \BibitemOpen
  \bibfield  {author} {\bibinfo {author} {\bibfnamefont {Y.}~\bibnamefont
  {Lanoisel\'ee}}, \bibinfo {author} {\bibfnamefont {N.}~\bibnamefont
  {Moutal}}, \ and\ \bibinfo {author} {\bibfnamefont {D.~S.}\ \bibnamefont
  {Grebenkov}},\ }\href@noop {} {\bibfield  {journal} {\bibinfo  {journal}
  {Nat. Comm.}\ }\textbf {\bibinfo {volume} {9}},\ \bibinfo {pages} {4398}
  (\bibinfo {year} {2018})}\BibitemShut {NoStop}%
\bibitem [{\citenamefont {Grebenkov}(2019)}]{grebenkov2019}%
  \BibitemOpen
  \bibfield  {author} {\bibinfo {author} {\bibfnamefont {D.~S.}\ \bibnamefont
  {Grebenkov}},\ }\href@noop {} {\bibfield  {journal} {\bibinfo  {journal} {J.
  Phys. A}\ }\textbf {\bibinfo {volume} {52}},\ \bibinfo {pages} {174001}
  (\bibinfo {year} {2019})}\BibitemShut {NoStop}%
\bibitem [{\citenamefont {Sposini}\ \emph {et~al.}(2019)\citenamefont
  {Sposini}, \citenamefont {Chechkin},\ and\ \citenamefont
  {Metzler}}]{sposini2019first}%
  \BibitemOpen
  \bibfield  {author} {\bibinfo {author} {\bibfnamefont {V.}~\bibnamefont
  {Sposini}}, \bibinfo {author} {\bibfnamefont {A.}~\bibnamefont {Chechkin}}, \
  and\ \bibinfo {author} {\bibfnamefont {R.}~\bibnamefont {Metzler}},\
  }\href@noop {} {\bibfield  {journal} {\bibinfo  {journal} {Journal of Physics
  A: Mathematical and Theoretical}\ }\textbf {\bibinfo {volume} {52}},\
  \bibinfo {pages} {04LT01} (\bibinfo {year} {2019})}\BibitemShut {NoStop}%
\bibitem [{\citenamefont {Sposini}\ \emph {et~al.}(2023)\citenamefont
  {Sposini}, \citenamefont {Nampoothiri}, \citenamefont {Chechkin},
  \citenamefont {Orlandini}, \citenamefont {Seno},\ and\ \citenamefont
  {Baldovin}}]{sposini2023pre}%
  \BibitemOpen
  \bibfield  {author} {\bibinfo {author} {\bibfnamefont {V.}~\bibnamefont
  {Sposini}}, \bibinfo {author} {\bibfnamefont {S.}~\bibnamefont
  {Nampoothiri}}, \bibinfo {author} {\bibfnamefont {A.}~\bibnamefont
  {Chechkin}}, \bibinfo {author} {\bibfnamefont {E.}~\bibnamefont {Orlandini}},
  \bibinfo {author} {\bibfnamefont {F.}~\bibnamefont {Seno}}, \ and\ \bibinfo
  {author} {\bibfnamefont {F.}~\bibnamefont {Baldovin}},\ }\href@noop {}
  {\bibfield  {journal} {\bibinfo  {journal} {Physical Review E}\ }\textbf
  {\bibinfo {volume} {109}},  \ \bibinfo {pages} {034120} (\bibinfo {year} {2024})}\BibitemShut
  {NoStop}%
\bibitem [{\citenamefont {Feller}(1968)}]{Feller1968}%
  \BibitemOpen
  \bibfield  {author} {\bibinfo {author} {\bibfnamefont {W.}~\bibnamefont
  {Feller}},\ }\href@noop {} {\emph {\bibinfo {title} {An Introduction to
  Probability Theory and Its Applications}}}\ (\bibinfo  {publisher} {John
  Wiley \& Sons},\ \bibinfo {year} {1968})\BibitemShut {NoStop}%
\bibitem [{\citenamefont {Bochner}(2020)}]{bochner2020harmonic}%
  \BibitemOpen
  \bibfield  {author} {\bibinfo {author} {\bibfnamefont {S.}~\bibnamefont
  {Bochner}},\ }\href@noop {} {\emph {\bibinfo {title} {Harmonic analysis and
  the theory of probability}}}\ (\bibinfo  {publisher} {University of
  California press},\ \bibinfo {year} {2020})\BibitemShut {NoStop}%
\bibitem [{\citenamefont {Meerson}\ and\ \citenamefont
  {Redner}(2015)}]{redner2015}%
  \BibitemOpen
  \bibfield  {author} {\bibinfo {author} {\bibfnamefont {B.}~\bibnamefont
  {Meerson}}\ and\ \bibinfo {author} {\bibfnamefont {S.}~\bibnamefont
  {Redner}},\ }\href@noop {} {\bibfield  {journal} {\bibinfo  {journal} {Phys.
  Rev. Lett.}\ }\textbf {\bibinfo {volume} {114}},\ \bibinfo {pages} {198101}
  (\bibinfo {year} {2015})}\BibitemShut {NoStop}%
\bibitem [{\citenamefont {Basnayake}\ \emph {et~al.}(2019)\citenamefont
  {Basnayake}, \citenamefont {Schuss},\ and\ \citenamefont
  {Holcman}}]{holcman2019a}%
  \BibitemOpen
  \bibfield  {author} {\bibinfo {author} {\bibfnamefont {K.}~\bibnamefont
  {Basnayake}}, \bibinfo {author} {\bibfnamefont {Z.}~\bibnamefont {Schuss}}, \
  and\ \bibinfo {author} {\bibfnamefont {D.}~\bibnamefont {Holcman}},\
  }\href@noop {} {\bibfield  {journal} {\bibinfo  {journal} {J. Nonlinear
  Sci.}\ }\textbf {\bibinfo {volume} {29}},\ \bibinfo {pages} {461} (\bibinfo
  {year} {2019})}\BibitemShut {NoStop}%
\bibitem [{\citenamefont {Schuss}\ \emph {et~al.}(2019)\citenamefont {Schuss},
  \citenamefont {Basnayake},\ and\ \citenamefont {Holcman}}]{holcman2019b}%
  \BibitemOpen
  \bibfield  {author} {\bibinfo {author} {\bibfnamefont {Z.}~\bibnamefont
  {Schuss}}, \bibinfo {author} {\bibfnamefont {K.}~\bibnamefont {Basnayake}}, \
  and\ \bibinfo {author} {\bibfnamefont {D.}~\bibnamefont {Holcman}},\
  }\href@noop {} {\bibfield  {journal} {\bibinfo  {journal} {Phys. Life Rev.}\
  }\textbf {\bibinfo {volume} {28}},\ \bibinfo {pages} {52} (\bibinfo {year}
  {2019})}\BibitemShut {NoStop}%
\bibitem [{\citenamefont {Redner}\ and\ \citenamefont
  {Meerson}(2019)}]{redner2019}%
  \BibitemOpen
  \bibfield  {author} {\bibinfo {author} {\bibfnamefont {S.}~\bibnamefont
  {Redner}}\ and\ \bibinfo {author} {\bibfnamefont {B.}~\bibnamefont
  {Meerson}},\ }\href@noop {} {\bibfield  {journal} {\bibinfo  {journal} {Phys.
  Life Rev.}\ }\textbf {\bibinfo {volume} {28}},\ \bibinfo {pages} {80}
  (\bibinfo {year} {2019})}\BibitemShut {NoStop}%
\bibitem [{\citenamefont {Martyushev}(2019)}]{martyushev2019}%
  \BibitemOpen
  \bibfield  {author} {\bibinfo {author} {\bibfnamefont {L.~M.}\ \bibnamefont
  {Martyushev}},\ }\href@noop {} {\bibfield  {journal} {\bibinfo  {journal}
  {Phys. Life Rev.}\ }\textbf {\bibinfo {volume} {28}},\ \bibinfo {pages} {83}
  (\bibinfo {year} {2019})}\BibitemShut {NoStop}%
\bibitem [{\citenamefont {Rusakov}\ and\ \citenamefont
  {Savtchenko}(2019)}]{rusakov2019}%
  \BibitemOpen
  \bibfield  {author} {\bibinfo {author} {\bibfnamefont {D.~A.}\ \bibnamefont
  {Rusakov}}\ and\ \bibinfo {author} {\bibfnamefont {L.~P.}\ \bibnamefont
  {Savtchenko}},\ }\href@noop {} {\bibfield  {journal} {\bibinfo  {journal}
  {Phys. Life Rev.}\ }\textbf {\bibinfo {volume} {28}},\ \bibinfo {pages} {85}
  (\bibinfo {year} {2019})}\BibitemShut {NoStop}%
\bibitem [{\citenamefont {Sokolov}(2019)}]{sokolov2019}%
  \BibitemOpen
  \bibfield  {author} {\bibinfo {author} {\bibfnamefont {I.~M.}\ \bibnamefont
  {Sokolov}},\ }\href@noop {} {\bibfield  {journal} {\bibinfo  {journal} {Phys.
  Life Rev.}\ }\textbf {\bibinfo {volume} {28}},\ \bibinfo {pages} {88}
  (\bibinfo {year} {2019})}\BibitemShut {NoStop}%
\bibitem [{\citenamefont {Coombs}(2019)}]{coombs2019}%
  \BibitemOpen
  \bibfield  {author} {\bibinfo {author} {\bibfnamefont {D.}~\bibnamefont
  {Coombs}},\ }\href@noop {} {\bibfield  {journal} {\bibinfo  {journal} {Phys.
  Life Rev.}\ }\textbf {\bibinfo {volume} {28}},\ \bibinfo {pages} {92}
  (\bibinfo {year} {2019})}\BibitemShut {NoStop}%
\bibitem [{\citenamefont {Tamm}(2019)}]{tamm2019}%
  \BibitemOpen
  \bibfield  {author} {\bibinfo {author} {\bibfnamefont {M.~V.}\ \bibnamefont
  {Tamm}},\ }\href@noop {} {\bibfield  {journal} {\bibinfo  {journal} {Phys.
  Life Rev.}\ }\textbf {\bibinfo {volume} {28}},\ \bibinfo {pages} {94}
  (\bibinfo {year} {2019})}\BibitemShut {NoStop}%
\bibitem [{\citenamefont {Basnayake}\ and\ \citenamefont
  {Holcman}(2019)}]{holcman2019c}%
  \BibitemOpen
  \bibfield  {author} {\bibinfo {author} {\bibfnamefont {K.}~\bibnamefont
  {Basnayake}}\ and\ \bibinfo {author} {\bibfnamefont {D.}~\bibnamefont
  {Holcman}},\ }\href@noop {} {\bibfield  {journal} {\bibinfo  {journal} {Phys.
  Life Rev.}\ }\textbf {\bibinfo {volume} {28}},\ \bibinfo {pages} {96}
  (\bibinfo {year} {2019})}\BibitemShut {NoStop}%
\bibitem [{\citenamefont {Lawley}(2020{\natexlab{a}})}]{lawley2020a}%
  \BibitemOpen
  \bibfield  {author} {\bibinfo {author} {\bibfnamefont {S.~D.}\ \bibnamefont
  {Lawley}},\ }\href@noop {} {\bibfield  {journal} {\bibinfo  {journal} {Phys.
  Rev. E}\ }\textbf {\bibinfo {volume} {101}},\ \bibinfo {pages} {012413}
  (\bibinfo {year} {2020}{\natexlab{a}})}\BibitemShut {NoStop}%
\bibitem [{\citenamefont {Lawley}(2020{\natexlab{b}})}]{lawley2020b}%
  \BibitemOpen
  \bibfield  {author} {\bibinfo {author} {\bibfnamefont {S.~D.}\ \bibnamefont
  {Lawley}},\ }\href@noop {} {\bibfield  {journal} {\bibinfo  {journal} {J.
  Math. Biol.}\ }\textbf {\bibinfo {volume} {80}},\ \bibinfo {pages} {2301}
  (\bibinfo {year} {2020}{\natexlab{b}})}\BibitemShut {NoStop}%
\bibitem [{\citenamefont {Chechkin}\ \emph {et~al.}(2017)\citenamefont
  {Chechkin}, \citenamefont {Seno}, \citenamefont {Metzler},\ and\
  \citenamefont {Sokolov}}]{chechkin2017}%
  \BibitemOpen
  \bibfield  {author} {\bibinfo {author} {\bibfnamefont {A.~V.}\ \bibnamefont
  {Chechkin}}, \bibinfo {author} {\bibfnamefont {F.}~\bibnamefont {Seno}},
  \bibinfo {author} {\bibfnamefont {R.}~\bibnamefont {Metzler}}, \ and\
  \bibinfo {author} {\bibfnamefont {I.~M.}\ \bibnamefont {Sokolov}},\
  }\href@noop {} {\bibfield  {journal} {\bibinfo  {journal} {Physical Review
  X}\ }\textbf {\bibinfo {volume} {7}},\ \bibinfo {pages} {021002} (\bibinfo
  {year} {2017})}\BibitemShut {NoStop}%
\bibitem [{\citenamefont {de~Gennes}(1979)}]{deGennes1979}%
  \BibitemOpen
  \bibfield  {author} {\bibinfo {author} {\bibfnamefont {P.-G.}\ \bibnamefont
  {de~Gennes}},\ }\href@noop {} {\emph {\bibinfo {title} {Scaling Concepts in
  Polymer Physics}}}\ (\bibinfo  {publisher} {Cornell University Press},\
  \bibinfo {year} {1979})\BibitemShut {NoStop}%
\bibitem [{\citenamefont {Vanderzande}(1998)}]{vanderzande1998}%
  \BibitemOpen
  \bibfield  {author} {\bibinfo {author} {\bibfnamefont {C.}~\bibnamefont
  {Vanderzande}},\ }\href@noop {} {\emph {\bibinfo {title} {Lattice Models of
  Polymers}}}\ (\bibinfo  {publisher} {Cambridge University Press},\ \bibinfo
  {year} {1998})\BibitemShut {NoStop}%
\bibitem [{\citenamefont {Flory}(1953)}]{flory1953}%
  \BibitemOpen
  \bibfield  {author} {\bibinfo {author} {\bibfnamefont {P.}~\bibnamefont
  {Flory}},\ }\href@noop {} {\emph {\bibinfo {title} {Principles of Polymer
  Chemistry}}}\ (\bibinfo  {publisher} {Cornell University Press},\ \bibinfo
  {year} {1953})\BibitemShut {NoStop}%
\bibitem [{\citenamefont {Yuste}\ and\ \citenamefont
  {Acedo}(2000)}]{yuste2000diffusion}%
  \BibitemOpen
  \bibfield  {author} {\bibinfo {author} {\bibfnamefont {S.}~\bibnamefont
  {Yuste}}\ and\ \bibinfo {author} {\bibfnamefont {L.}~\bibnamefont {Acedo}},\
  }\href@noop {} {\bibfield  {journal} {\bibinfo  {journal} {Journal of Physics
  A: Mathematical and General}\ }\textbf {\bibinfo {volume} {33}},\ \bibinfo
  {pages} {507} (\bibinfo {year} {2000})}\BibitemShut {NoStop}%
\bibitem [{\citenamefont {Holcman}\ and\ \citenamefont
  {Schuss}(2015)}]{holcman2015stochastic}%
  \BibitemOpen
  \bibfield  {author} {\bibinfo {author} {\bibfnamefont {D.}~\bibnamefont
  {Holcman}}\ and\ \bibinfo {author} {\bibfnamefont {Z.}~\bibnamefont
  {Schuss}},\ }\href@noop {} {\emph {\bibinfo {title} {Stochastic Narrow Escape in
Molecular and Cellular Biology. Analysis and Applications}}} (\bibinfo  {publisher} {Springer, New York},\ \bibinfo
  {year} {2015})\BibitemShut
  {NoStop}%
\bibitem [{\citenamefont {Weiss}\ \emph {et~al.}(1983)\citenamefont {Weiss},
  \citenamefont {Shuler},\ and\ \citenamefont {Lindenberg}}]{weiss1983}%
  \BibitemOpen
  \bibfield  {author} {\bibinfo {author} {\bibfnamefont {G.}~\bibnamefont
  {Weiss}}, \bibinfo {author} {\bibfnamefont {K.}~\bibnamefont {Shuler}}, \
  and\ \bibinfo {author} {\bibfnamefont {K.}~\bibnamefont {Lindenberg}},\
  }\href@noop {} {\bibfield  {journal} {\bibinfo  {journal} {J. Stat. Phys.}\
  }\textbf {\bibinfo {volume} {31}},\ \bibinfo {pages} {255} (\bibinfo {year}
  {1983})}\BibitemShut {NoStop}%
\bibitem [{\citenamefont {Mej\'{\i}a-Monasterio}\ \emph
  {et~al.}(2011)\citenamefont {Mej\'{\i}a-Monasterio}, \citenamefont
  {Oshanin},\ and\ \citenamefont {Schehr}}]{oshanin2011}%
  \BibitemOpen
  \bibfield  {author} {\bibinfo {author} {\bibfnamefont {C.}~\bibnamefont
  {Mej\'{\i}a-Monasterio}}, \bibinfo {author} {\bibfnamefont {G.}~\bibnamefont
  {Oshanin}}, \ and\ \bibinfo {author} {\bibfnamefont {G.}~\bibnamefont
  {Schehr}},\ }\href@noop {} {\bibfield  {journal} {\bibinfo  {journal} {J.
  Stat. Mech.}\ ,\ \bibinfo {pages} {P06022}} (\bibinfo {year}
  {2011})}\BibitemShut {NoStop}%
\bibitem [{\citenamefont {Corless}\ \emph {et~al.}(1996)\citenamefont
  {Corless}, \citenamefont {Gonnet}, \citenamefont {Hare}, \citenamefont
  {Jeffrey},\ and\ \citenamefont {Knuth}}]{lambertW}%
  \BibitemOpen
  \bibfield  {author} {\bibinfo {author} {\bibfnamefont {R.}~\bibnamefont
  {Corless}}, \bibinfo {author} {\bibfnamefont {G.}~\bibnamefont {Gonnet}},
  \bibinfo {author} {\bibfnamefont {D.}~\bibnamefont {Hare}}, \bibinfo {author}
  {\bibfnamefont {D.}~\bibnamefont {Jeffrey}}, \ and\ \bibinfo {author}
  {\bibfnamefont {D.}~\bibnamefont {Knuth}},\ }\href@noop {} {\bibfield
  {journal} {\bibinfo  {journal} {Adv. Comput. Math.}\ }\textbf {\bibinfo
  {volume} {5}},\ \bibinfo {pages} {329} (\bibinfo {year} {1996})}\BibitemShut
  {NoStop}%
\bibitem [{\citenamefont {Nampoothiri}\ \emph {et~al.}(2021)\citenamefont
  {Nampoothiri}, \citenamefont {Orlandini}, \citenamefont {Seno},\ and\
  \citenamefont {Baldovin}}]{nampoothiri2021}%
  \BibitemOpen
  \bibfield  {author} {\bibinfo {author} {\bibfnamefont {S.}~\bibnamefont
  {Nampoothiri}}, \bibinfo {author} {\bibfnamefont {E.}~\bibnamefont
  {Orlandini}}, \bibinfo {author} {\bibfnamefont {F.}~\bibnamefont {Seno}}, \
  and\ \bibinfo {author} {\bibfnamefont {F.}~\bibnamefont {Baldovin}},\
  }\href@noop {} {\bibfield  {journal} {\bibinfo  {journal} {Physical Review
  E}\ }\textbf {\bibinfo {volume} {104}},\ \bibinfo {pages} {L062501} (\bibinfo
  {year} {2021})}\BibitemShut {NoStop}%
\bibitem [{\citenamefont {Nampoothiri}\ \emph {et~al.}(2022)\citenamefont
  {Nampoothiri}, \citenamefont {Orlandini}, \citenamefont {Seno},\ and\
  \citenamefont {Baldovin}}]{nampoothiri2022}%
  \BibitemOpen
  \bibfield  {author} {\bibinfo {author} {\bibfnamefont {S.}~\bibnamefont
  {Nampoothiri}}, \bibinfo {author} {\bibfnamefont {E.}~\bibnamefont
  {Orlandini}}, \bibinfo {author} {\bibfnamefont {F.}~\bibnamefont {Seno}}, \
  and\ \bibinfo {author} {\bibfnamefont {F.}~\bibnamefont {Baldovin}},\
  }\href@noop {} {\bibfield  {journal} {\bibinfo  {journal} {New J. Phys.}\
  }\textbf {\bibinfo {volume} {24}},\ \bibinfo {pages} {023003} (\bibinfo
  {year} {2022})}\BibitemShut {NoStop}%
\bibitem [{\citenamefont {Marcone}\ \emph {et~al.}(2022)\citenamefont
  {Marcone}, \citenamefont {Nampoothiri}, \citenamefont {Orlandini},
  \citenamefont {Seno},\ and\ \citenamefont {Baldovin}}]{marcone2022}%
  \BibitemOpen
  \bibfield  {author} {\bibinfo {author} {\bibfnamefont {B.}~\bibnamefont
  {Marcone}}, \bibinfo {author} {\bibfnamefont {S.}~\bibnamefont
  {Nampoothiri}}, \bibinfo {author} {\bibfnamefont {E.}~\bibnamefont
  {Orlandini}}, \bibinfo {author} {\bibfnamefont {F.}~\bibnamefont {Seno}}, \
  and\ \bibinfo {author} {\bibfnamefont {F.}~\bibnamefont {Baldovin}},\
  }\href@noop {} {\bibfield  {journal} {\bibinfo  {journal} {J. Phys. A: Math.
  Theor.}\ }\textbf {\bibinfo {volume} {55}},\ \bibinfo {pages} {354003}
  (\bibinfo {year} {2022})}\BibitemShut {NoStop}%
\bibitem [{\citenamefont {Heston}(1993)}]{heston1993}%
  \BibitemOpen
  \bibfield  {author} {\bibinfo {author} {\bibfnamefont {S.}~\bibnamefont
  {Heston}},\ }\href@noop {} {\bibfield  {journal} {\bibinfo  {journal} {Rev.
  Financial Studies}\ }\textbf {\bibinfo {volume} {6}},\ \bibinfo {pages} {327}
  (\bibinfo {year} {1993})}\BibitemShut {NoStop}%
\bibitem [{\citenamefont {Fouqu\'e}\ \emph {et~al.}(2000)\citenamefont
  {Fouqu\'e}, \citenamefont {Papanicolaou},\ and\ \citenamefont
  {Sircar}}]{sircar2000}%
  \BibitemOpen
  \bibfield  {author} {\bibinfo {author} {\bibfnamefont {J.-P.}\ \bibnamefont
  {Fouqu\'e}}, \bibinfo {author} {\bibfnamefont {G.}~\bibnamefont
  {Papanicolaou}}, \ and\ \bibinfo {author} {\bibfnamefont {K.}~\bibnamefont
  {Sircar}},\ }\href@noop {} {\emph {\bibinfo {title} {Derivatives in Financial
  Markets with Stochastic Volatility}}}\ (\bibinfo  {publisher} {Cambridge
  University Press, Cambridge, England},\ \bibinfo {year} {2000})\BibitemShut
  {NoStop}%
\bibitem [{\citenamefont {Gillespie}(1992)}]{gillespie1992}%
  \BibitemOpen
  \bibfield  {author} {\bibinfo {author} {\bibfnamefont {D.~T.}\ \bibnamefont
  {Gillespie}},\ }\href@noop {} {\emph {\bibinfo {title} {Markov Processes: An
  Introduction for Physical Scientists}}}\ (\bibinfo  {publisher} {San Diego,
  CA: Academic Press},\ \bibinfo {year} {1992})\BibitemShut {NoStop}%
\bibitem [{\citenamefont {Doi}\ and\ \citenamefont {F}(1992)}]{Doi1992}%
  \BibitemOpen
  \bibfield  {author} {\bibinfo {author} {\bibfnamefont {M.}~\bibnamefont
  {Doi}}\ and\ \bibinfo {author} {\bibfnamefont {E.~S.}\ \bibnamefont {F}},\
  }\href@noop {} {\emph {\bibinfo {title} {The Theory of Polymer Dynamics}}}\
  (\bibinfo  {publisher} {Oxford University Press},\ \bibinfo {year}
  {1992})\BibitemShut {NoStop}%
\bibitem [{\citenamefont {Beck}\ and\ \citenamefont {Cohen}(2003)}]{beck2003}%
  \BibitemOpen
  \bibfield  {author} {\bibinfo {author} {\bibfnamefont {C.}~\bibnamefont
  {Beck}}\ and\ \bibinfo {author} {\bibfnamefont {E.~G.}\ \bibnamefont
  {Cohen}},\ }\href@noop {} {\bibfield  {journal} {\bibinfo  {journal} {Physica
  A: Statistical mechanics and its applications}\ }\textbf {\bibinfo {volume}
  {322}},\ \bibinfo {pages} {267} (\bibinfo {year} {2003})}\BibitemShut
  {NoStop}%
\bibitem [{\citenamefont {Beck}(2006)}]{beck2006}%
  \BibitemOpen
  \bibfield  {author} {\bibinfo {author} {\bibfnamefont {C.}~\bibnamefont
  {Beck}},\ }\href@noop {} {\bibfield  {journal} {\bibinfo  {journal} {Progress
  of Theoretical Physics Supplement}\ }\textbf {\bibinfo {volume} {162}},\
  \bibinfo {pages} {29} (\bibinfo {year} {2006})}\BibitemShut {NoStop}%
\bibitem [{\citenamefont {Touchette}(2009)}]{touchette2009}%
  \BibitemOpen
  \bibfield  {author} {\bibinfo {author} {\bibfnamefont {H.}~\bibnamefont
  {Touchette}},\ }\href@noop {} {\bibfield  {journal} {\bibinfo  {journal}
  {Physics Reports}\ }\textbf {\bibinfo {volume} {478}},\ \bibinfo {pages} {1}
  (\bibinfo {year} {2009})}\BibitemShut {NoStop}%
\bibitem [{\citenamefont {Odian}(2004)}]{odian2004}%
  \BibitemOpen
  \bibfield  {author} {\bibinfo {author} {\bibfnamefont {G.}~\bibnamefont
  {Odian}},\ }\href@noop {} {\emph {\bibinfo {title} {Principles of
  Polymerization}}}\ (\bibinfo  {publisher} {John Wiley \& Sons},\ \bibinfo
  {year} {2004})\BibitemShut {NoStop}%
\bibitem [{\citenamefont {Sposini}\ \emph {et~al.}(2018)\citenamefont
  {Sposini}, \citenamefont {Chechkin}, \citenamefont {Seno}, \citenamefont
  {Pagnini},\ and\ \citenamefont {Metzler}}]{sposini2018}%
  \BibitemOpen
  \bibfield  {author} {\bibinfo {author} {\bibfnamefont {V.}~\bibnamefont
  {Sposini}}, \bibinfo {author} {\bibfnamefont {A.~V.}\ \bibnamefont
  {Chechkin}}, \bibinfo {author} {\bibfnamefont {F.}~\bibnamefont {Seno}},
  \bibinfo {author} {\bibfnamefont {G.}~\bibnamefont {Pagnini}}, \ and\
  \bibinfo {author} {\bibfnamefont {R.}~\bibnamefont {Metzler}},\ }\href@noop
  {} {\bibfield  {journal} {\bibinfo  {journal} {New Journal of Physics}\
  }\textbf {\bibinfo {volume} {20}},\ \bibinfo {pages} {043044} (\bibinfo
  {year} {2018})}\BibitemShut {NoStop}%
\bibitem [{\citenamefont {Pacheco-Pozo}\ and\ \citenamefont
  {Sokolov}(2021)}]{sokolov2021}%
  \BibitemOpen
  \bibfield  {author} {\bibinfo {author} {\bibfnamefont {A.}~\bibnamefont
  {Pacheco-Pozo}}\ and\ \bibinfo {author} {\bibfnamefont {I.~M.}\ \bibnamefont
  {Sokolov}},\ }\href@noop {} {\bibfield  {journal} {\bibinfo  {journal}
  {Physical Review Letters}\ }\textbf {\bibinfo {volume} {127}},\ \bibinfo
  {pages} {120601} (\bibinfo {year} {2021})}\BibitemShut {NoStop}%
\bibitem [{\citenamefont {Wang}\ \emph {et~al.}(2020)\citenamefont {Wang},
  \citenamefont {Seno}, \citenamefont {Sokolov}, \citenamefont {Chechkin},\
  and\ \citenamefont {Metzler}}]{wang2020unexpected}%
  \BibitemOpen
  \bibfield  {author} {\bibinfo {author} {\bibfnamefont {W.}~\bibnamefont
  {Wang}}, \bibinfo {author} {\bibfnamefont {F.}~\bibnamefont {Seno}}, \bibinfo
  {author} {\bibfnamefont {I.~M.}\ \bibnamefont {Sokolov}}, \bibinfo {author}
  {\bibfnamefont {A.~V.}\ \bibnamefont {Chechkin}}, \ and\ \bibinfo {author}
  {\bibfnamefont {R.}~\bibnamefont {Metzler}},\ }\href@noop {} {\bibfield
  {journal} {\bibinfo  {journal} {New Journal of Physics}\ }\textbf {\bibinfo
  {volume} {22}},\ \bibinfo {pages} {083041} (\bibinfo {year}
  {2020})}\BibitemShut {NoStop}%
\bibitem [{\citenamefont {Sadoon}\ and\ \citenamefont
  {Wang}(2018)}]{sadoon2018anomalous}%
  \BibitemOpen
  \bibfield  {author} {\bibinfo {author} {\bibfnamefont {A.~A.}\ \bibnamefont
  {Sadoon}}\ and\ \bibinfo {author} {\bibfnamefont {Y.}~\bibnamefont {Wang}},\
  }\href@noop {} {\bibfield  {journal} {\bibinfo  {journal} {Physical Review
  E}\ }\textbf {\bibinfo {volume} {98}},\ \bibinfo {pages} {042411} (\bibinfo
  {year} {2018})}\BibitemShut {NoStop}%
\bibitem [{\citenamefont {Cherstvy}\ \emph {et~al.}(2019)\citenamefont
  {Cherstvy}, \citenamefont {Thapa}, \citenamefont {Wagner},\ and\
  \citenamefont {Metzler}}]{cherstvy2019non}%
  \BibitemOpen
  \bibfield  {author} {\bibinfo {author} {\bibfnamefont {A.~G.}\ \bibnamefont
  {Cherstvy}}, \bibinfo {author} {\bibfnamefont {S.}~\bibnamefont {Thapa}},
  \bibinfo {author} {\bibfnamefont {C.~E.}\ \bibnamefont {Wagner}}, \ and\
  \bibinfo {author} {\bibfnamefont {R.}~\bibnamefont {Metzler}},\ }\href@noop
  {} {\bibfield  {journal} {\bibinfo  {journal} {Soft Matter}\ }\textbf
  {\bibinfo {volume} {15}},\ \bibinfo {pages} {2526} (\bibinfo {year}
  {2019})}\BibitemShut {NoStop}%
\bibitem [{\citenamefont {Sabri}\ \emph {et~al.}(2020)\citenamefont {Sabri},
  \citenamefont {Xu}, \citenamefont {Krapf},\ and\ \citenamefont
  {Weiss}}]{sabri2020elucidating}%
  \BibitemOpen
  \bibfield  {author} {\bibinfo {author} {\bibfnamefont {A.}~\bibnamefont
  {Sabri}}, \bibinfo {author} {\bibfnamefont {X.}~\bibnamefont {Xu}}, \bibinfo
  {author} {\bibfnamefont {D.}~\bibnamefont {Krapf}}, \ and\ \bibinfo {author}
  {\bibfnamefont {M.}~\bibnamefont {Weiss}},\ }\href@noop {} {\bibfield
  {journal} {\bibinfo  {journal} {Physical Review Letters}\ }\textbf {\bibinfo
  {volume} {125}},\ \bibinfo {pages} {058101} (\bibinfo {year}
  {2020})}\BibitemShut {NoStop}%
\bibitem [{\citenamefont {Han}\ \emph {et~al.}(2020)\citenamefont {Han},
  \citenamefont {Korabel}, \citenamefont {Chen}, \citenamefont {Johnston},
  \citenamefont {Gavrilova}, \citenamefont {Allan}, \citenamefont {Fedotov},\
  and\ \citenamefont {Waigh}}]{han2020deciphering}%
  \BibitemOpen
  \bibfield  {author} {\bibinfo {author} {\bibfnamefont {D.}~\bibnamefont
  {Han}}, \bibinfo {author} {\bibfnamefont {N.}~\bibnamefont {Korabel}},
  \bibinfo {author} {\bibfnamefont {R.}~\bibnamefont {Chen}}, \bibinfo {author}
  {\bibfnamefont {M.}~\bibnamefont {Johnston}}, \bibinfo {author}
  {\bibfnamefont {A.}~\bibnamefont {Gavrilova}}, \bibinfo {author}
  {\bibfnamefont {V.~J.}\ \bibnamefont {Allan}}, \bibinfo {author}
  {\bibfnamefont {S.}~\bibnamefont {Fedotov}}, \ and\ \bibinfo {author}
  {\bibfnamefont {T.~A.}\ \bibnamefont {Waigh}},\ }\href@noop {} {\bibfield
  {journal} {\bibinfo  {journal} {Elife}\ }\textbf {\bibinfo {volume} {9}},\
  \bibinfo {pages} {e52224} (\bibinfo {year} {2020})}\BibitemShut {NoStop}%
\bibitem [{\citenamefont {Benelli}\ and\ \citenamefont
  {Weiss}(2021)}]{benelli2021sub}%
  \BibitemOpen
  \bibfield  {author} {\bibinfo {author} {\bibfnamefont {R.}~\bibnamefont
  {Benelli}}\ and\ \bibinfo {author} {\bibfnamefont {M.}~\bibnamefont
  {Weiss}},\ }\href@noop {} {\bibfield  {journal} {\bibinfo  {journal} {New
  Journal of Physics}\ }\textbf {\bibinfo {volume} {23}},\ \bibinfo {pages}
  {063072} (\bibinfo {year} {2021})}\BibitemShut {NoStop}%
\bibitem [{\citenamefont {Janczura}\ \emph {et~al.}(2021)\citenamefont
  {Janczura}, \citenamefont {Balcerek}, \citenamefont {Burnecki}, \citenamefont
  {Sabri}, \citenamefont {Weiss},\ and\ \citenamefont
  {Krapf}}]{janczura2021identifying}%
  \BibitemOpen
  \bibfield  {author} {\bibinfo {author} {\bibfnamefont {J.}~\bibnamefont
  {Janczura}}, \bibinfo {author} {\bibfnamefont {M.}~\bibnamefont {Balcerek}},
  \bibinfo {author} {\bibfnamefont {K.}~\bibnamefont {Burnecki}}, \bibinfo
  {author} {\bibfnamefont {A.}~\bibnamefont {Sabri}}, \bibinfo {author}
  {\bibfnamefont {M.}~\bibnamefont {Weiss}}, \ and\ \bibinfo {author}
  {\bibfnamefont {D.}~\bibnamefont {Krapf}},\ }\href@noop {} {\bibfield
  {journal} {\bibinfo  {journal} {New Journal of Physics}\ }\textbf {\bibinfo
  {volume} {23}},\ \bibinfo {pages} {053018} (\bibinfo {year}
  {2021})}\BibitemShut {NoStop}%
\bibitem [{\citenamefont {Korabel}\ \emph {et~al.}(2021)\citenamefont
  {Korabel}, \citenamefont {Han}, \citenamefont {Taloni}, \citenamefont
  {Pagnini}, \citenamefont {Fedotov}, \citenamefont {Allan},\ and\
  \citenamefont {Waigh}}]{korabel2021local}%
  \BibitemOpen
  \bibfield  {author} {\bibinfo {author} {\bibfnamefont {N.}~\bibnamefont
  {Korabel}}, \bibinfo {author} {\bibfnamefont {D.}~\bibnamefont {Han}},
  \bibinfo {author} {\bibfnamefont {A.}~\bibnamefont {Taloni}}, \bibinfo
  {author} {\bibfnamefont {G.}~\bibnamefont {Pagnini}}, \bibinfo {author}
  {\bibfnamefont {S.}~\bibnamefont {Fedotov}}, \bibinfo {author} {\bibfnamefont
  {V.}~\bibnamefont {Allan}}, \ and\ \bibinfo {author} {\bibfnamefont {T.~A.}\
  \bibnamefont {Waigh}},\ }\href@noop {} {\bibfield  {journal} {\bibinfo
  {journal} {Entropy}\ }\textbf {\bibinfo {volume} {23}},\ \bibinfo {pages}
  {958} (\bibinfo {year} {2021})}\BibitemShut {NoStop}%
\bibitem [{\citenamefont {Speckner}\ and\ \citenamefont
  {Weiss}(2021)}]{speckner2021single}%
  \BibitemOpen
  \bibfield  {author} {\bibinfo {author} {\bibfnamefont {K.}~\bibnamefont
  {Speckner}}\ and\ \bibinfo {author} {\bibfnamefont {M.}~\bibnamefont
  {Weiss}},\ }\href@noop {} {\bibfield  {journal} {\bibinfo  {journal}
  {Entropy}\ }\textbf {\bibinfo {volume} {23}},\ \bibinfo {pages} {892}
  (\bibinfo {year} {2021})}\BibitemShut {NoStop}%
\bibitem [{\citenamefont {Balcerek}\ \emph {et~al.}(2022)\citenamefont
  {Balcerek}, \citenamefont {Burnecki}, \citenamefont {Thapa}, \citenamefont
  {Wy{\l}oma{\'n}ska},\ and\ \citenamefont
  {Chechkin}}]{balcerek2022fractional}%
  \BibitemOpen
  \bibfield  {author} {\bibinfo {author} {\bibfnamefont {M.}~\bibnamefont
  {Balcerek}}, \bibinfo {author} {\bibfnamefont {K.}~\bibnamefont {Burnecki}},
  \bibinfo {author} {\bibfnamefont {S.}~\bibnamefont {Thapa}}, \bibinfo
  {author} {\bibfnamefont {A.}~\bibnamefont {Wy{\l}oma{\'n}ska}}, \ and\
  \bibinfo {author} {\bibfnamefont {A.}~\bibnamefont {Chechkin}},\ }\href@noop
  {} {\bibfield  {journal} {\bibinfo  {journal} {Chaos: An Interdisciplinary
  Journal of Nonlinear Science}\ }\textbf {\bibinfo {volume} {32}}, 
  \ \bibinfo {pages} {093114} (\bibinfo
  {year} {2022})}\BibitemShut {NoStop}%
\bibitem [{\citenamefont {Wang}\ \emph {et~al.}(2023)\citenamefont {Wang},
  \citenamefont {Balcerek}, \citenamefont {Burnecki}, \citenamefont {Chechkin},
  \citenamefont {Janusonis}, \citenamefont {Slezak}, \citenamefont {Vojta},
  \citenamefont {Wylomanska},\ and\ \citenamefont {Metzler}}]{wang2023memory}%
  \BibitemOpen
  \bibfield  {author} {\bibinfo {author} {\bibfnamefont {W.}~\bibnamefont
  {Wang}}, \bibinfo {author} {\bibfnamefont {M.}~\bibnamefont {Balcerek}},
  \bibinfo {author} {\bibfnamefont {K.}~\bibnamefont {Burnecki}}, \bibinfo
  {author} {\bibfnamefont {A.~V.}\ \bibnamefont {Chechkin}}, \bibinfo {author}
  {\bibfnamefont {S.}~\bibnamefont {Janusonis}}, \bibinfo {author}
  {\bibfnamefont {J.}~\bibnamefont {Slezak}}, \bibinfo {author} {\bibfnamefont
  {T.}~\bibnamefont {Vojta}}, \bibinfo {author} {\bibfnamefont
  {A.}~\bibnamefont {Wylomanska}}, \ and\ \bibinfo {author} {\bibfnamefont
  {R.}~\bibnamefont {Metzler}},\ }\href@noop {} {\bibfield  {journal} {\bibinfo
   {journal} {arXiv preprint arXiv:2303.01551}\ } (\bibinfo {year}
  {2023})}\BibitemShut {NoStop}%
\bibitem [{\citenamefont {Korabel}\ \emph {et~al.}(2023)\citenamefont
  {Korabel}, \citenamefont {Taloni}, \citenamefont {Pagnini}, \citenamefont
  {Allan}, \citenamefont {Fedotov},\ and\ \citenamefont
  {Waigh}}]{korabel2023ensemble}%
  \BibitemOpen
  \bibfield  {author} {\bibinfo {author} {\bibfnamefont {N.}~\bibnamefont
  {Korabel}}, \bibinfo {author} {\bibfnamefont {A.}~\bibnamefont {Taloni}},
  \bibinfo {author} {\bibfnamefont {G.}~\bibnamefont {Pagnini}}, \bibinfo
  {author} {\bibfnamefont {V.}~\bibnamefont {Allan}}, \bibinfo {author}
  {\bibfnamefont {S.}~\bibnamefont {Fedotov}}, \ and\ \bibinfo {author}
  {\bibfnamefont {T.~A.}\ \bibnamefont {Waigh}},\ }\href@noop {} {\bibfield
  {journal} {\bibinfo  {journal} {Scientific Reports}\ }\textbf {\bibinfo
  {volume} {13}},\ \bibinfo {pages} {8789} (\bibinfo {year}
  {2023})}\BibitemShut {NoStop}%
\end{thebibliography}
\end{document}